\documentclass[twocolumn,prd,nofootinbib,superscriptaddress]{revtex4-1}
\usepackage{color}
\usepackage{amsmath}
\usepackage{mathrsfs}
\usepackage{graphicx}
\usepackage{booktabs}
\usepackage{bm}
\usepackage{makecell}
\usepackage{xcolor}

\usepackage{adjustbox} 

\usepackage[bookmarksnumbered,plainpages,linkbordercolor={0 1 1}, pdfborder={0 1 1},pdfauthor={Your Name},pdftitle={Title},pdfsubject={Paper}, pdfkeywords={keywords}]{hyperref}

\newcommand{\optional}[1]{}

\newcommand{\scD}{\mathcal{D}}
\newcommand{\scY}{\mathcal{Y}}
\newcommand{\scE}{\mathcal{E}}
\newcommand{\scL}{\mathcal{L}}
\newcommand{\scN}{\mathcal{N}}
\newcommand{\scC}{\mathcal{C}}
\newcommand{\scS}{\mathcal{S}}
\newcommand{\scV}{\mathcal{V}}
\newcommand{\Msun}{M_{\odot}}
\newcommand{\consf}[1]{{\fontfamily{lmtt}\selectfont{#1}}}
\newcommand{\hyperpipe}{\consf{HyperPipe}}

\def\RIT{Center for Computational Relativity and Gravitation, Rochester Institute of Technology, Rochester, New York
	14623, USA}

\begin{document}
	
\title{Improved Population and EOS Joint Inference for Binary Neutron Star Systems}

\author{Marc Ebiri}
\affiliation{\RIT}
\author{Richard O'Shaughnessy}
\affiliation{\RIT}

\begin{abstract}
	\noindent The extreme environment within neutron stars presents the opportunity to probe the nuclear equation of state at high densities while studying the properties of these stellar remnants. Historically, the equation of state and neutron star mass distribution have been inferred separately, with the latter often simply assumed to be fixed in studies of the former, but the dependence of the equation of state on neutron star mass indicates they should be inferred together. 
	In this work, we extend the generalized version of the popular RIFT algorithm known as \hyperpipe{} to interface with flexible, user-provided priors that will facilitate joint inference of equation of state and binary neutron star population hyperparameters. We demonstrate this framework's utility via application to a combination of several existing observations, particularly including massive galactic pulsars, double neutron star systems, millisecond X-ray pulsars, the gravitational wave event GW170817 and the nuclear symmetry energy, within the context of a widely-adopted parametric EOS family and a simple Gaussian population model. We recover parameters consistent with previous analyses of the EOS and binary neutron star population, finding the latter to fit a bivariate normal distribution with mean $(\mu_1,\mu_2) = (1.39,1.27) \Msun$ and width $\sigma = 0.08 \Msun$. Furthermore, we implement novel coordinate transformations in our pipeline, with which we have discovered that the ad hoc prior boundaries used for this EOS family may be too restrictive. We present results with relaxed yet still physical prior boundaries, noting vastly improved hyperparameter posteriors and modestly different equation of state inferences. 
\end{abstract}

\maketitle

\section{Introduction}

The nuclear equation of state (EOS) - the relationship between pressure and density in cold nuclear matter - remains weakly constrained by terrestrial experiments, with differences having substantial impact on the predicted properties of neutron stars \cite{Lattimer_2016,Baym_2018}. Astrophysical observations of neutron star systems, however, provide a natural mechanism by which to investigate the nuclear EOS, and countless studies have therefore sought to constrain it via measurements of neutron stars (NS), improving and refining their estimates as more observations have become available; e.g., \cite{GW170817_2019,Abbott_2018,Carney_2018,Dietrich_2020,Hebeler_2013,Landry_2020,Miller_2020,askold2024}. For example, the size of isolated, rotating neutron stars is encoded in the pulsed X-ray emission from their surface, allowing observations and theoretical modeling of galactic X-ray sources to limit the range of possible neutron star mass-radius relationships \cite{Ozel_2010,Lattimer_2014, Miller_2019,Bogdanov_2019}. Meanwhile, neutron stars in coalescing binaries are subject to strong tidal interactions in the late stages of inspiral, which have an observable impact on the outgoing gravitational wave signal \cite{Flanagan_2008,Favata_2014,Dietrich2019a} and thus enable constraints on the nuclear EOS \cite{DelPozzo2013,Agathos_2015,Lackey_2015}. With GW170817, the imprint of these tidal interactions on the inspiral signal was first constrained \cite{GW170817_2019,GW170817paper,De_2018,Abbott_2020}, affording widely-investigated implications for the nuclear equation of state (e.g., \cite{Abbott_2018}).

Previous studies (e.g., \cite{Landry_2020,atul2025,Dietrich_2020}) have sought to constrain the nuclear EOS by joining together distinct measurements of NS properties, but many of these analyses chose to assume the distribution of the corresponding NS population rather than infer it, often selecting a uniform distribution or, rarely, a more complex model (e.g., \cite{dan2020}). However, this approach introduces biases into the inference \cite{Wysocki_2019}, since there exists an inherent relationship between a neutron star's equation of state and its mass encoded in its tidal deformability \cite{Wade_2014}. It is therefore logical to estimate these properties simultaneously. 



In this work we extend the general-purpose code pipeline called \consf{HyperPipe} introduced by Kedia et al. \cite{atul2025}, itself a generalization of the RIFT inference engine \cite{rift2018}, to implement a flexible method for simultaneously inferring the binary NS population and the nuclear EOS. Similar recent studies \cite{anik2025,Golomb_2025,Ghosh_2025} have pursued this goal using distinct methodologies; our approach largely parallels that of \cite{dan2020}, except that we choose a simpler population model, since our focus is primarily on implementing backend code that can support those more sophisticated models in future studies. In addition, we implement two other significant improvements to \consf{HyperPipe}, adopting a superior coordinate system for fitting our hyperparameter posteriors and altering how our code explores the hyperparameter space to increase our efficiency. While doing so, we also investigate the suitability of the prior bounds typically employed for our EOS model. 

We organize the present work as follows. Section \ref{sec:methods} details our inference framework and priors, Section \ref{sec:data} outlines the data sources we incorporate and their corresponding likelihood factors, while Section \ref{sec:hyperpipe} overviews our inference pipeline and our modifications to it. Section \ref{sec:results} presents and discusses our results, and Section \ref{sec:conclusion} concludes with a note on how they can be further improved with new data in the future. 

\section{Joint Inference Framework}
\label{sec:methods}

Bayesian inference expresses the posterior distribution $p(\scY|d)$ as a likelihood $\mathcal{L}(d|\scY)$ times a prior $p(\scY)$ and a normalization factor $C$ \cite{Thrane_2019}:
\begin{equation}\label{eq:bayes}
	p(\scY|d) = C\,\mathcal{L}(d|\scY)\,p(\scY)
\end{equation}
\noindent where $C$ is called the ``evidence'' and can be defined as:
\begin{equation}\label{eq:norm}
	\frac{1}{C} \equiv \int d\scY \scL(d|\scY) p(\scY)
\end{equation}

\noindent Our goal, then, is to estimate $p(\scY | d)$ via the \consf{HyperPipe} algorithm, which takes in the data $d$ to compute Eqns. \ref{eq:bayes} and \ref{eq:norm} simultaneously via iterative calculations of $\scL(d|\scY) p(\scY)$ (see Sec. \ref{sec:hyperpipe}). Furthermore, since independent observations of an event possess independent statistical uncertainties and are associated with distinct astrophysical phenomena, we can combine multiple observations together via the product of their individual likelihoods. Thus, for a set of hyperparameters $\scY$ with a prior $p(\scY)$, we can expand the likelihood in Eq. (\ref{eq:bayes}) as a product of likelihood functions of the form:
\begin{equation}\label{eq:lnet}
\scL_{\rm net}(\scY) = \left[\prod_{k}\scD_k(\scY)\right]\hspace{-.15cm}\left[\prod_{g}\scE_g(\scY)\right]\hspace{-.15cm}\left[\prod_{n}\mathcal{Z}_n(\scY)\right]\hspace{-.15cm}\,\scC(\scY)\,\scS(\scY)
\end{equation}
\noindent In this equation, $k$ indexes DNS mass observations, $g$ indexes gravitational wave observations, and $n$ indexes x-ray neutron star observations. We detail each of these factors and their corresponding data sources in Sec. \ref{sec:data}. First, however, we describe our prior models for the EOS and population hyperparameters, since all of these factors depend in some fashion on assumptions about the NS populations responsible for the associated events, which we encode in the prior model $p(\scY)$. For brevity, hereafter we use a subscript $\alpha$ to refer to an expression conditioned on a specific EOS realization; i.e., $\scY = \{\scY_{\alpha}\}$.

\subsection{EOS framework}\label{sec:eos}

The structure of neutron stars in a static, spherically symmetric limit is governed by the TOV equations \cite{Opp_Volkoff,Tolman}. Solving these equations requires a mass-radius relation $m(r)$, dependent on the baryon density $\rho$, and a pressure-density relation $P(\rho)$: the equation of state.

Following previous work (e.g., \cite{Carney_2018, Abbott_2018, Miller_2019, dan2020}), we employ the spectral EOS parameterization introduced by Lindblom \cite{Lindblom_2010, Lindblom_2018}, which approximates the zero-temperature adiabatic index $\Gamma(p)$ as a 4-term power series expansion in the basis of a dimensionless pressure variable $x=\log(p/p_0)$:
\begin{equation}\label{eq:gamma}
\Gamma(p) = \exp\left[\sum_{k=0}^{3} \gamma_k\left(\log p/p_0\right)^k\right]
\end{equation}
\noindent The reference pressure $p_0$ is often (e.g., \cite{Miller_2021}) taken to be the pressure at half the nuclear saturation density, $p_0 = n_{sat}/2$, in accordance with the calculation in \cite{Hebeler_2013} that it is approximately the crust-core transition density. 
While we do not explicitly enforce other typical constraints (e.g., bounding the adiabatic index) on this model, our TOV solver only allows equations of state that are nearly causal, so any highly unphysical samples are rejected by default in our algorithm.

The hyperparameter ranges we employ for these $\gamma_k$ coordinates, which were introduced in \cite{Carney_2018} and have become customary for EOS inference using GW data, are shown in the top row of Table \ref{tbl:bounds} below. 
Unfortunately, as noted in \cite{dan2020}, the only EOS realizations produced by the spectral decomposition model that are physically allowed lie in a narrow, strongly-correlated subspace $\scV$ of the conventional four-dimensional Cartesian space $\scC$ with coordinates $\gamma_k$, occupying a region only $\scV \simeq 0.005\%$ of $\scC$. 
Because the axes of $\scC$ do not align with the axes that would define the subspace $\scV$, sampling along them is massively inefficient: for example, for a posterior grid created on the first iteration of a \hyperpipe{} analysis using Eq. (\ref{eq:lnet}), only $\sim 1$\% of the samples drawn were physical. 


In this work, we endeavor to mitigate this issue by introducing two new strategies in our sampling algorithm: a rotated coordinate system, described below, and a reflection mechanism, described in Sec. \ref{sec:hyperpipe}. As a result, we employ the traditional $\gamma_k$ parameter ranges only for the comparative analysis discussed in Appendix \ref{sec:app}. 

\subsubsection{Rotated Coordinate System}

Sampling efficiency can be substantially increased without altering the (local) prior density by using an affine transformation, choosing the coordinate system of 
a more efficicent hypercube $\scC'$ that closely bounds the shape of $\scV$, such that uniform sampling produces a much higher fraction of physical EOS realizations. Before fitting and sampling our posteriors, then, we rotate the EOS parameter sets $\scY$, which initially lie in the hypercube $\scC$ with axes ($\gamma_0$, $\gamma_1$, $\gamma_2$, $\gamma_3$), to the new hypercube $\scC'$ with axes ($r'_0$, $r'_1$, $r'_2$, $r'_3$) using the transformation described by Wysocki et al. \cite{dan2020}. We apply the inverse of this rotation to the sampled posteriors in $r'_i$ coordinates to obtain our final posteriors in $\gamma_k$ coordinates. 

\begin{table*}[t!] 
	\caption{Hypercube bounds used in our analyses in both the Cartesian $\gamma_i$ space $\scC$ and in the rotated $r'_i$ space $\scC'$, with the latter shown with buffered bounds of 10\% and 400\%.} \label{tbl:bounds}
	\begin{center}
		\begin{tabular}{l cccc}
			\hline
			Hypercube Space & $\gamma_0$ & $\gamma_1$ & $\gamma_2$ & $\gamma_3$ \\
			\hline & & & & \vspace{-.3cm}\\
			$\vspace{.05cm}\scC$ & [+0.2, +2.0] & [$-1.6$, +1.7] & [$-0.6$, +0.6] & [$-0.02$, +0.02] \\ 
			\hline & & & & \vspace{-.3cm}\\
			& $r'_0$ & $r'_1$ & $r'_2$ & $r'_3$ \\
			\hline & & & & \vspace{-.3cm}\\
			\vspace{.05cm}$\scC'$ & [$-4.37722$, +4.91227] & [$-1.82240$, +2.06387] & [$-0.32445$, +0.36469] & [$-0.09529$, +0.11426] \\ 
			\vspace{.05cm}$\scC'$ +10\% & [$-4.814942$, +5.403497] & [$-2.004640$, +2.270257] & [$-0.356895$, +0.401159] & [$-0.104819$, +0.121506] \\ 
			\vspace{.05cm}$\scC'$ +400\% & [$-21.88610$, +24.56135] & [$-9.11200$, +10.31935] & [$-1.62225$, +1.82345] & [$-0.47645$, +0.55230] \\ 
			\hline & & & & \vspace{-.5cm}
		\end{tabular}
	\end{center}
\end{table*}

For our hyperparameter ranges in the $r'_i$ coordinates, we adopt the bounds found by \cite{dan2020}, reproduced in the second row of Table \ref{tbl:bounds}. However, note that, due to the necessarily limited sample size used to estimate $\scV$, the hypercube $\scC'$ formed from these bounds may not enclose all of $\scV$. To accomodate this possibility, we simply add a fractional buffer to each of the $r'_i$ bounds; while this slightly reduces our sampling efficiency, it should ensure that we allow the full space of physical EOS hyperparameters to be sampled. 

To check this expectation, we perform two analyses in this work that are identical except for the size of their buffer factor. Our primary analysis uses the 10\% buffer suggested by \cite{dan2020}, while our additional test vastly expands the prior ranges with a buffer of 400\%. The exact values of these bounds are presented in the bottom two rows of Table \ref{tbl:bounds}. Although the 10\% buffer is expected to encompass all of the physical subspace for valid EOS hyperparameters, the larger buffer aims to provide a clearer view of how the hyperparameter posteriors behave when sampling in another coordinate system.

\subsubsection{Initial EOS Samples}

For our initial injected EOS samples, we utilize a publicly available datafile for GW170817 from the LIGO DCC \cite{170817EOSdata}, which contains posterior samples for the 4 $\Gamma$-spectral EOS hyperparameters $\gamma_k$. All our initial parameter sets $\{\scY_{\alpha}\}$ therefore produce a valid EOS model.

\subsection{Population model}\label{sec:pop}

Efforts to characterize the NS population date back several decades (e.g., \cite{Finn_1994}), and over time two general approaches have developed. In one approach, neutron stars in different types of systems are treated as entirely separate populations \cite{Ozel_2012,Kiziltan_2013,atul2025,anik2025,Landry_2021,Golomb_2025}, while in the other, all NS sources are inferred as part of one universal population \cite{dan2020,Antoniadis_2016,Alsing_2018,Golomb_2022}. The latter method requires careful attention towards population selection effects, since neutron stars in different species of systems (NS-WD, DNS, LMXB, etc.) can follow different evolutionary paths \cite{anik2025} and indeed seem to fit distinct population distributions \cite{Landry_2021,Golomb_2025}. We focus only on the binary neutron star (BNS/DNS) population in this work, and thus the mass measurements we incorporate for our population inference, described in Sec. \ref{sec:dns}, only belong to those systems. 

In contrast to previous works employing similar likelihood evaluation techniques, which assumed uniform priors \cite{atul2025, Landry_2020} or priors dependent on other parameters \cite{Miller_2021, Miller_2020} for their inference, we assume an independent, non-evolving (\cite{anik2025}) and non-uniform distribution for the BNS population in gravitational mass $m$. Specifically, we model the population as a truncated multivariate normal distribution, with limits $(m_1,m_2) \in [1.0 M_{\odot}, 3.0 M_{\odot}]$ and $m_1 \ge m_2$. We therefore extend our set of prior parameters $\scY$ to include two population means $\mu_1$ and $\mu_2$, subject to fiducial bounds $1.0 M_{\odot}\le \mu_2 \le \mu_1 \le 3.0 M_{\odot}$, and a single uncertainty parameter $\sigma$ that applies to both means, such that we can infer the BNS population simultaneously with the EOS within our algorithm. 

Note that this prior constitutes an assumption about the astrophysical distribution of real neutron stars, which need not reach the maximum mass theoretically allowed by the EOS. Additionally, a fully self-consistent population model needs to account for parameter-dependent selection effects \cite{Wysocki_2019, Mandel_2019}, but to simplify the discussion for our present analysis we neglect these.

\begin{figure}
	\centering
	\includegraphics[width=.45\textwidth]{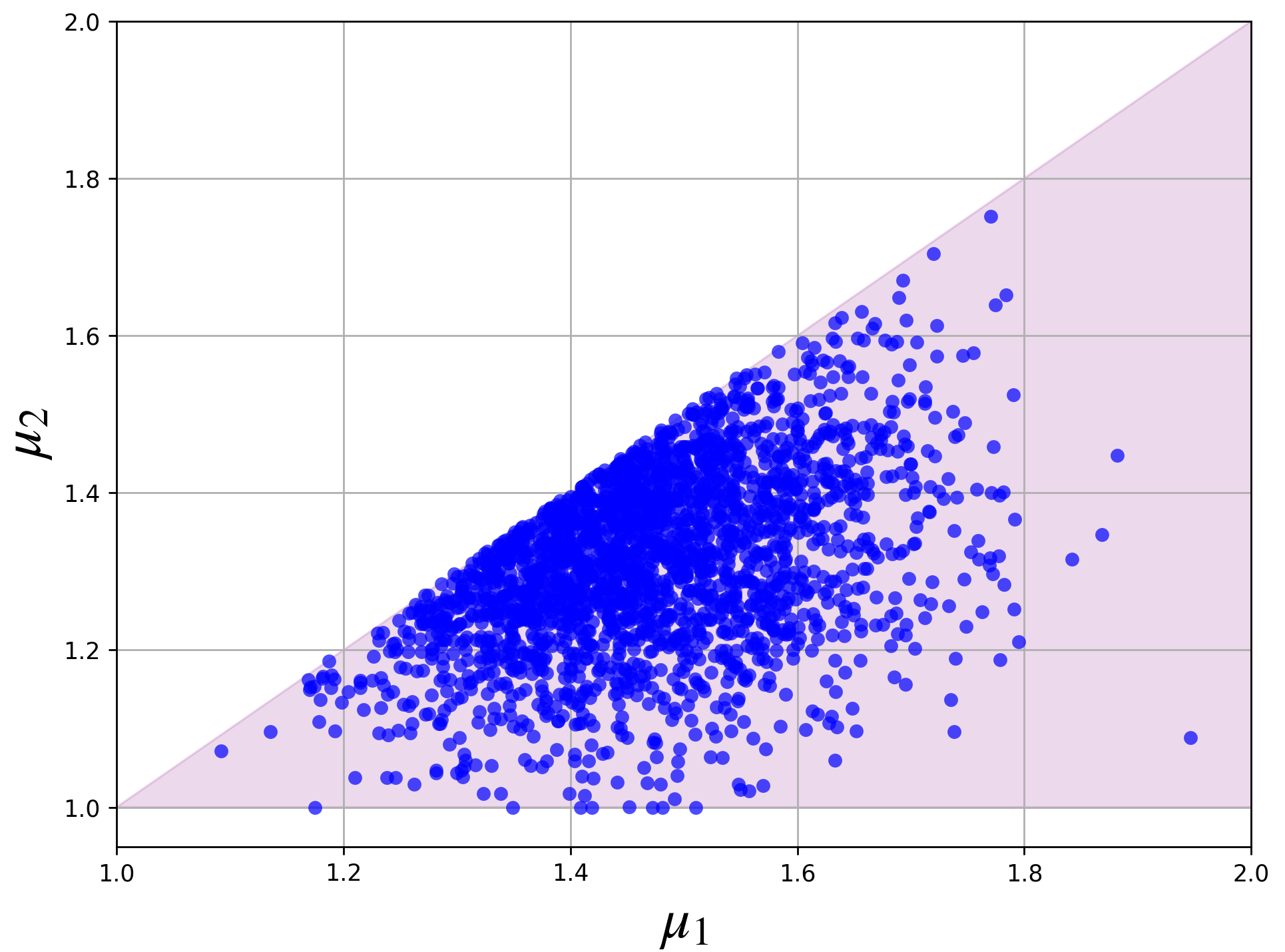}
	\caption{Initial distribution of population samples ($\mu_1$, $\mu_2$), drawn from a multivariate normal with mean $(\mu_1,\mu_2) = (1.39, 1.39) \Msun$. All population samples throughout the inference procedure in this study are truncated within the shaded triangular region $\mu_2 \ge 1$, $\mu_1 \le 3$, $\mu_1 \ge \mu_2$. \label{fig:popgrid}}
\end{figure}

We generate an initial sample set for the new prior parameters $(\mu_1,\mu_2,\sigma)$ by sampling points from two normal distributions. We draw the means $(\mu_1,\mu_2)$, sorted to ensure $\mu_1 \ge \mu_2$, from a two-dimensional distribution with mean values $\mu_{1,pop} = \mu_{2,pop} = 1.39 \Msun$ and width $\sigma_{pop} = 0.14 \Msun$ in both dimensions, and we draw the widths $\sigma$ from a one-dimensional distribution with mean $\sigma_{pop}$ and standard deviation $\sigma_{pop}/4$, requiring only that $\sigma > 0$. We obtain our values for $(\mu_{i,pop}, \sigma_{pop})$ from the DNS population parameter results of Anik et al. \cite{anik2025} for their modified linear EOS model; while these values were thus produced via a different EOS model than our own, we use them only as a starting point, much as our initial EOS samples are drawn from previous GW170817 results. This initial population distribution is plotted in Fig. \ref{fig:popgrid} to illustrate the region in which our model is bounded.
We append these sets of parameters $\{(\mu_1,\mu_2,\sigma)\}$ to the EOS samples described in the preceding subsection to form the initial proposed parameter sets $\scY_0$ for our inference pipeline. 

In addition to the EOS and masses, we must also assume a prior on the NS spin. For our analyses, we employ the usual prior for the aligned spin $z$-component introduced by \cite{rift2018}, which distributes the spin isotropically in a sphere and uniformly in magnitude, enforcing a maximum dimensionless value $a = S/m^2 < 0.05$ \cite{atul2025}. This prior becomes equivalent to that for uniform spin magnitude after marginalizing out other degrees of freedom.

\section{Data Factors and Sources}\label{sec:data}

We next describe the data and equations utilized in computing each of the factors present in Eq. (\ref{eq:lnet}).

\subsection{Galactic DNS Masses}\label{sec:dns}

The majority of known NS masses have been measured via the binary interactions of galactic radio pulsars, which account for only about 250 of over 3000 known pulsars \cite{Ozel_2016}. Radio measurements of pulsars in binary systems obtain mass estimates by accounting for a mixture of Keplerian and post-Keplerian effects, most notably the Shapiro time delay (e.g., \cite{Cromartie_2019_J0740}). These methods have been used to identify neutron stars across a range of masses up to 2.35 $\Msun$ \cite{Romani_2022}. 

\begin{figure}[t]
	\centering
	\includegraphics[width=.48\textwidth]{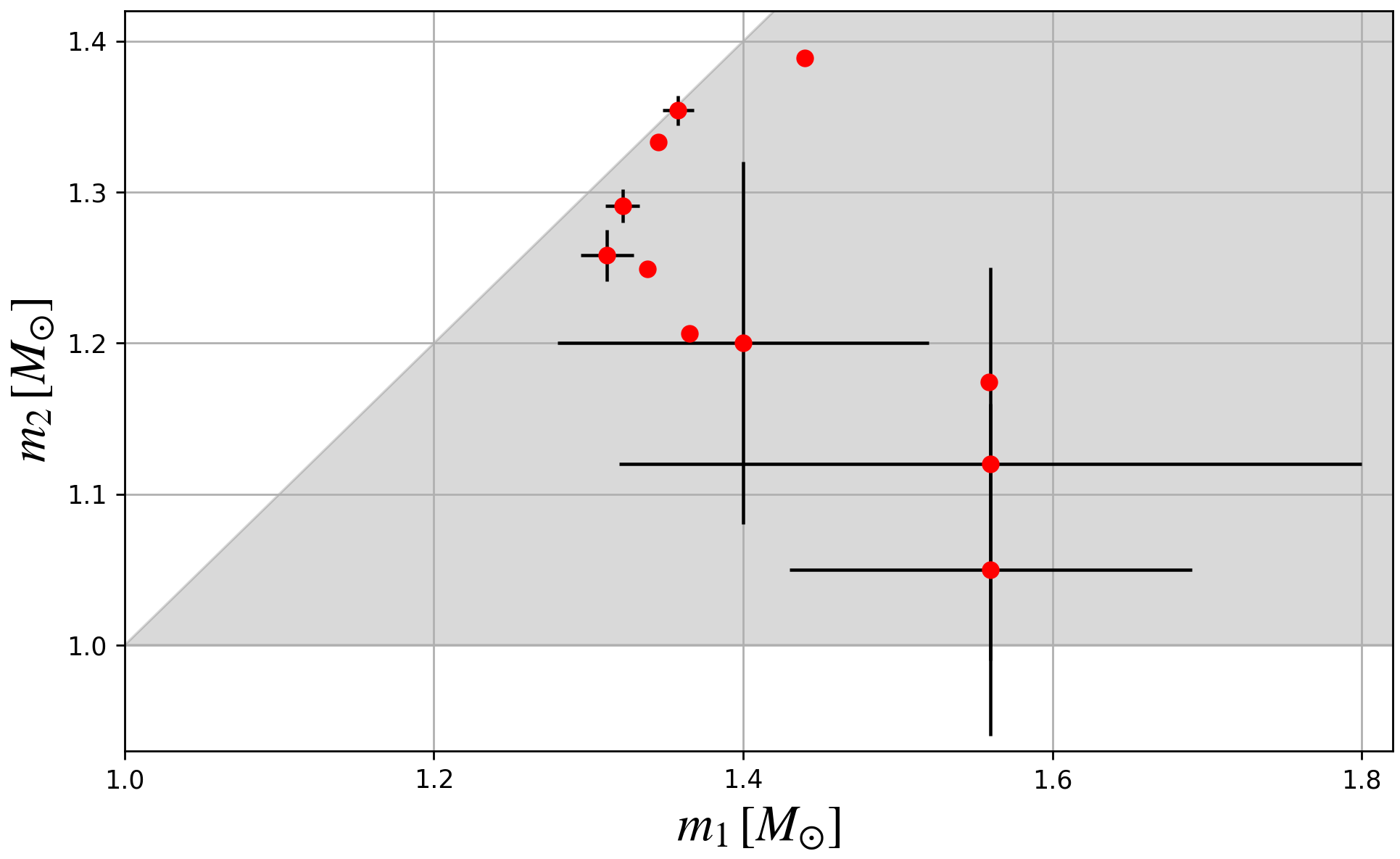}
	\caption{Masses of 11 DNS systems, plotted with their 1-$\sigma$ errorbars. Several masses have been measured to such high precision that their errorbars are not visible. The shaded area indictates the region to which we constrain the population masses, with bounds $1 \Msun \le m_2 \le m_1 \le 3 \Msun$. \label{fig:dns}}
\end{figure}

For our analysis, we use 11 pairs of DNS masses compiled by \cite{anik2025}, reproduced in Table \ref{tbl:dns} and plotted in Fig. \ref{fig:dns}, to weight the likelihood that any sample drawn from our population distribution is part of the real DNS population. For the $k$-th real mass pair $m_k=(m_{1,k}, m_{2,k})$, this weighting is accomplished by integrating over the product of two-dimensional normal distributions for the sample mass and real mass, as:
\begin{equation}\label{eq:dk}
\scD_k = \iint \scN(m | \mu_{\alpha}, \sigma_{\alpha})\,g_k(m|m_k,\sigma_k) d^2m
\end{equation}
\noindent This integral is performed over a cut-corner-rectangular region, bounded by $m_k-3\sigma_k \le m \le m_k+3\sigma_k$ in both dimensions where those bounds do not exceed our fixed population boundaries of $1.0 \Msun \le m \le 3.0 \Msun$ and $m_1 = m_2$. The narrow measurement uncertainties $\sigma_k$ on several of these masses cause this factor to strongly downweight any parameter set with masses even slightly distant from this measured population sample.

\begin{table}[t]
	\caption{Masses and 1-$\sigma$ uncertainties of 11 pairs of DNS systems, organized such that $m_1 > m_2$ \cite{anik2025}.} 
	\label{tbl:dns}
	\begin{center}
		\begin{tabular}{l l l l l}
			\hline
			System & $m_1\,[\Msun]$ & $m_2\,[\Msun]$ & $\sigma_1\,[\Msun]$ & $\sigma_2\,[\Msun]$ \\
			\hline
			J0453+1559 & 1.559 & 1.174 & 0.004 & 0.004\\
			J1906+0746 & 1.322 &1.291 &0.011& 0.011\\
			B1534+12   & 1.3452 &1.3332& 0.001& 0.001\\
			B1913+16   & 1.4398 &1.3886& 0.0002& 0.0002\\
			B2127+11C  & 1.358 &1.354 &0.01& 0.01\\
			J0737-3039AB& 1.3381& 1.2489& 0.0007& 0.0007\\
			J1756-2251 &1.312 &1.258 &0.017& 0.017\\
			J1807-2500B &1.3655& 1.2064& 0.0021& 0.002\\
			J1518+4904 &1.56 &1.05 &0.13& 0.11\\
			J1811-1736 &1.56 &1.12 &0.24& 0.13\\
			J1829+2456 &1.40 &1.20 &0.12& 0.12\\ 
			\hline 
		\end{tabular}
	\end{center}
\end{table}

\subsection{Pulsar Maximum Mass}\label{sec:mmax}

For each EOS there is a maximum possible mass allowed by general relativity \cite{Watts_2016,Chamel_2013}. As a result, the EOS can be indirectly probed using very massive pulsars, whose measured masses bound the maximum mass from below and rule out EOS models that are not stiff enough to support such massive stars.

We follow \cite{Dietrich_2020} in expressing the likelihood of an EOS consistent with the observations of three massive pulsars, shown in Table \ref{tbl:mmax}, as the product of three normal distributions:
\begin{equation}\label{eq:cdf}
\scC_{\alpha} = \prod_i\Phi\left(\frac{M_{max,\alpha} - m_i}{\sigma_i}\right)
\end{equation}
\noindent where $\Phi(z)$ is the cumulative distribution function (CDF). These pulsar observations suggest that the maximum NS mass is larger than 2.1$\Msun$, and we therefore apply the product $\scC_{\alpha}$ as a constraint factor in Eq. (\ref{eq:lnet}), reducing the likelihood of any EOS parameter set $\scY_{\alpha}$ that supports a maximum mass lower than these observed masses. The masses of these pulsars are shown as horizontal bands in the mass-radius plot Fig. \ref{fig:nicer} below. 

\begin{table}
	\caption{Masses and 1-$\sigma$ uncertainties of three of the most massive pulsars known, as in \cite{Dietrich_2020}. Note that we use the newer value for J1614 from \cite{nanograv_2018}, while our value for J0740 has been superseded by that in \cite{Fonseca_2021}. \label{tbl:mmax}}
	\begin{center}
		\begin{tabular}{l l l}
			\hline
			Pulsar & $M\, [\Msun]$ & $\sigma\, [\Msun]$ \\
			\hline
			PSR J0740+6620 & 2.14 & 0.1 \\ 
			PSR J0348+0432 & 2.01 & 0.04 \\ 
			PSR J1614-2230 & 1.908 & 0.016 \\ 
			\hline 
		\end{tabular}
	\end{center}
\end{table}

We can also impose a fixed lower bound on the maximum mass, automatically discarding any EOS with $M_{max} < 1.8 \Msun$ as unphysical: since we have several well-measured neutron star masses above this bound, there is no point investigating EOS parameters that cannot support them. While it is common practice for physical EOS model development to require $M_{max} > 2.0 \Msun$ \cite{Watts_2016}, we retain our more generous bound, both to acknowledge the arguments made by \cite{Miller_2020} (especially since we use data for PSR J1614, with $M < 2 \Msun$) and to let our algorithm reach this conclusion independently via the strong downweighting provided by Eq. (\ref{eq:cdf}). As a result, we find that our posteriors are not significantly affected by which of these cutoff values we employ.

\subsection{Gravitational Waves}\label{sec:gw}

Neutron stars in coalescing binaries are subject to strong tidal interactions in the late stages of inspiral, which have an observable impact on the outgoing gravitational wave signal \cite{Flanagan_2008, Favata_2014} and thus enable constraints on the nuclear EOS \cite{DelPozzo2013, Agathos_2015, Lackey_2015}. During the late stages of the inspiral of coalescing neutron stars, the gravitational gradient across the diameter of one binary component due to its companion's field induces a quadrupole moment in the star, causing enhanced emission of gravitational radiation and a slight acceleration to the inspiral phase. The magnitude of the induced quadrupole moment is related to neutron stars' internal structure, with larger stars being less compact and thus more easily deformable under the influence of an external field. This effect is quantified through the dimensionless tidal deformability $\Lambda$, defined as the ratio of the induced quadrupole moment to external perturbing tidal field \cite{Flanagan_2008}:

\begin{equation}\label{eq:lambdam}
\Lambda = \frac{2}{3}k_2\left(\frac{c^2r}{Gm}\right)^5
\end{equation} 

\noindent where $k_2$ is the $l = 2$ Love number, $m$ is the mass, and $r$ is the radius. As this equation indicates, the utility of the tidal deformability for EOS inference lies in its uniqueness: there is only one value $\Lambda(m|\rm EOS)$ for each given set of EOS parameters and mass. Since the individual $\Lambda_i$ for each NS in the binary are strongly correlated and thus difficult to distinguish within an observed GW signal \cite{Wade_2014}, the leading order contribution to the phase evolution of a GW inspiral is typically measured via the weighted combination of $\Lambda$ terms 

\begin{equation}\label{eq:lambdatilde}
\tilde{\Lambda} = \frac{16}{13}\frac{(m_1 + 12m_2)m_1^4\Lambda_1 + (m_2 + 12m_1)m_2^4\Lambda_2}{(m_1 + m_2)^5}
\end{equation}

We assume this contribution fully characterizes the effect the EOS has on emitted gravitational waves. Following the procedure of \cite{atul2025}, we utilize the RIFT inference algorithm \cite{Pankow_2015,rift2018,expanding_rift2023,narrowing_rift2025} to generate marginal likelihoods $\scL_{marg}(X)$ for many candidate points $X = (m_1,m_2,\boldsymbol{\chi}_1, \boldsymbol{\chi}_2,\Lambda_1,\Lambda_2)$ via the waveform model IMRPhenomPv2\_NRTidalv2 \cite{Dietrich2019a, Dietrich2019b}. This model incorporates precession physics while omitting higher-order modes, but we still allow only for aligned $z$-component spins in our analysis, such that $\boldsymbol{\chi}_i = (0, 0, \chi_{i,z})$, using the $z$-component prior mentioned in Sec. \ref{sec:pop}. RIFT's iterative assessments produce estimates of the posterior $p_{\rm post}$ for the intrinsic parameters $X$, proportional to the product $\scL(X)p_{gw}(X)$ of a fiducial population prior $p_{gw}(X)$ and an interpolated estimate $\hat{\scL}(X)$ of the true likelihood $\scL(X)$, as: 

\begin{equation}\label{eq:riftpost}
p_{\rm post} = \frac{\scL(X)p_{gw}(X)}{\int dX\, \scL(X)p_{gw}(X)},
\end{equation}

\noindent These estimates $\scL_{marg}(X)$ are computed by RIFT's CIP (Construct Intrinsic Posterior) stage via random forest interpolation and exported as a data product containing \{($X, \scL_{marg}(X)$)\}. We then input this data into \consf{HyperPipe}, interpolating it via the EOS-derived relationship connecting the tides $\Lambda(m|\scY)$ to the NS mass $m$ and EOS model $\scY$. The result is a new, EOS-informed marginal likelihood $\scL_{marg}(X)$ with which we compute the marginal evidence: 
\begin{equation}\label{eq:cip}
\scE = \int \scL_{marg}(X(x,\scY)) p(x)\,dx
\end{equation}
\noindent Here, $x = (m_1, m_2, \chi_{1,z}, \chi_{2,z})$ are the intrinsic binary parameters, which $X$ supplements with the EOS-derived tidal parameters: $X(x,\scY) = (\mathcal{M}_c, \eta, \chi_{1,z}, \chi_{2,z}, \Lambda(m_1|\scY),\Lambda(m_2|\scY))$, where $\mathcal{M}_c$ is the chirp mass, $\mathcal{M}_c = (m_1m_2)^{3/5}/(m_1+m_2)^{1/5}$, and $\eta = m_1m_2/(m_1+m_2)^2$ is the symmetric mass ratio. The $p(x)$ term is the prior on the parameters in $x$, which is uniform for all except the NS population parameters $m_1$, $m_2$, to which we apply our model described in Sec. \ref{sec:pop}.

Note that Eq. (\ref{eq:lnet}) indicates that multiple GW events $g$ can be incorporated into the overall likelihood factor $\prod_g \scE_g(\scY)$, and indeed other studies have considered multiple dark BNS events \cite{Ghosh_2025} or have included the signal GW190425 (e.g., \cite{Landry_2020}), which originated from a more massive system that may or may not have been a BNS merger and, correspondingly, had weaker tidal interactions as well as an overall low SNR signal \cite{GW190425}. For simplicity, we utilize only data for GW170817 in this work; thus, $g = 1$ here.

\subsection{X-ray Pulsars}\label{sec:nicer}

Beginning in 2017, the Neutron Star Interior Composition Explorer (NICER) experiment has observed periodic thermal soft X-ray emission from hot spots on the surfaces of millisecond pulsars (MSPs) in both isolated and binary systems \cite{NICER_2016}. Time-series observations of this emission from an MSP accumulates a pulse profile, which can be fit using soft X-ray waveform models and estimates for the hot spots' size, shape, and number. Determining these properties allows the NS mass and radius to be jointly inferred \cite{Ozel_2016,Watts_2016, Bogdanov_2019}. Significantly, in contrast to older estimates that were susceptible to glitches and systematic errors \cite{Miller_2019,Ozel_2010}, the high SNR of the NICER data increases the precision of the measured mass and radius, leading to tighter constraints on the inferred EOS. 

\begin{figure}[t]
	\centering
	\includegraphics[width=.49\textwidth]{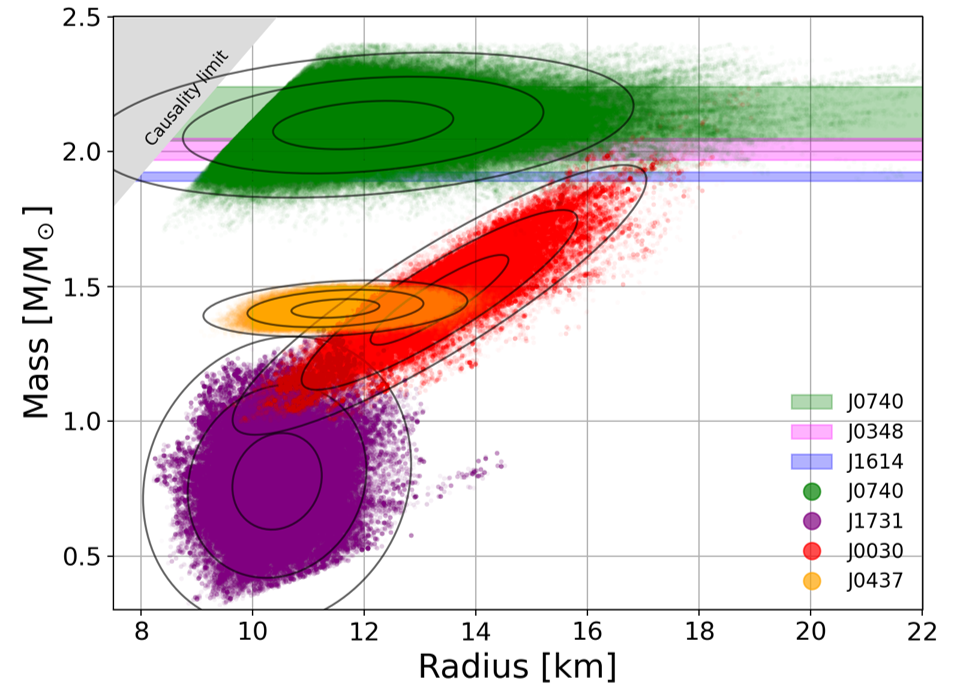}
	\caption{Mass vs. radius posterior samples of NICER pulse profiles for three millisecond pulsars, PSR-J0030, PSR-J0437, and PSR-J0740, and the HESS-J1731 object. Overlaid are ellipses representing the two-dimensional normal distributions with which we approximate the data in our calculations. We also show the mass bounds for the three high-mass pulsars listed in Table \ref{tbl:mmax}. \label{fig:nicer}}
\end{figure}

We utilize data from observations of three NICER objects in this work: PSR J0030+0451 \cite{J0030data,Miller_2019}, PSR J0740+6620 \cite{J0740data,Miller_2021}, and PSR J0437-4715 \cite{J0437data,Choudhury_2024}. We also include the HESS source J1731-347 \cite{J1731data,Doroshenko_2022}. This is a larger composition of data than other studies (e.g., \cite{atul2025,Landry_2020,Mendes_2026}) have used. For HESS J1731, we parallel \cite{atul2025,J1731new} in using only the x-ray data from \cite{Doroshenko_2022} that assumes a uniformly-emitting, single-temperature carbon atmosphere model and Gaia parallax priors. The weighted mass-radius posterior samples reported for each data set are shown in Fig. \ref{fig:nicer} to give an idea of their shape.  

For our analysis, we simply adopt all the conventions of \cite{atul2025}, approximating the posteriors for each NICER object as two-dimensional normal distributions (shown as overlaid ellipses in Fig. \ref{fig:nicer}), and we compute for these Gaussian likelihoods $\scL(m,r)$ the marginal likelihood for each EOS $\scY_{\alpha}$ using the same kind of evidence integral as Eq. (\ref{eq:norm}):
\begin{equation}\label{eq:nicerinit}
	\mathcal{Z}(\scY_{\alpha}) = \int \scL(m_{\alpha}(s),r_{\alpha}(s))\,p(s)\, ds
\end{equation}
\noindent where $s$ is a fiducial parameter characterizing the dimensionless
prior $p(s)$ along each one-parameter family in the EoS $\scY_{\alpha}$. Unlike for the GW likelihood in Eq. (\ref{eq:cip}), however, we do not implement our Gaussian population model from Sec. \ref{sec:pop} as the prior here; instead we adopt a uniform mass prior with fixed, EOS-independent limits $m \in [m_{min}, m_{max}] = [0.4 \Msun, 2.1 \Msun]$. Our population model is for binary neutron star masses, while NICER sources are primarily isolated pulsars \cite{Watts_2016}, and we have no reason to assume that these populations share the same distribution. We do, however, assume the same distribution from \cite{rift2018} for neutron star spin. We therefore choose $p(s)$ to be compatible with this uniform mass prior and rewrite Eq. (\ref{eq:nicerinit}) to depend on $m$:
\begin{equation}\label{eq:nicer}
\mathcal{Z}(\scY_{\alpha}) = \mathcal{F} \int \scL(m(s), r_{\alpha}(m(s)))\,dm
\end{equation}
\noindent where $\mathcal{F} = 1/(m_{max} - m_{min})$ is the necessary normalization constant. To compute the radius $R_{\alpha}(m(s))$ for a given $\scY_{\alpha}$ and mass $m$ drawn from our uniform distribution, we utilize the \consf{RePrimAnd} library developed by Wolfgang Kastaun \cite{Kastaun_2021, reprimand}, which provides a robust TOV solver to find stable branch solutions and recover primitive variables, like pressure and baryon rest-mass density, for a given set of EOS parameters $\scY_{\alpha}$. 

\subsection{Nuclear Symmetry Energy}\label{sec:symenergy}

\noindent Finally, we incorporate an additional constraining factor on nuclear-density matter. At the nuclear saturation number density $n_{sat}$, the symmetry energy $S$ can be defined as the difference between the energies per baryon of pure nuclear matter (PNM) and symmetric nuclear matter (SNM):
\begin{equation}
	S(n_{sat}) \simeq \frac{\epsilon_{\mathrm{PNM}}(n_{sat})}{n_{sat}}- \frac{\epsilon_{\mathrm{SNM}}(n_{sat})}{n_{sat}}
\end{equation}
\noindent Following \cite{atul2025,Mroczek_2023}, we fix the saturation density and binding energy, rather than propagating their uncertainties. We set $n_{sat}=0.16\,\mathrm{fm}^{-3}$, which corresponds to the rest-mass density $\rho_{sat} \simeq 2.7\times 10^{14}\,\mathrm{g\,cm}^{-3}$ \cite{Tsang_2012, Li_2019}, while we fix the SNM binding energy per baryon at saturation to $B=-16$ MeV and take $m_n c^2=939.6$ MeV \cite{Mroczek_2023}. We then evaluate the symmetry-energy proxy
\begin{align}
	S_{\alpha}^{\mathrm{proxy}}(n_{sat}) &=
	\frac{\epsilon_{\alpha}(n_{sat})}{n_{sat}}-
	\left(m_n c^2+B\right)\nonumber\\
	&=\frac{\epsilon_{\alpha}(n_{sat})}{n_{sat}}-923.6\,\mathrm{MeV},
\end{align}
\noindent where $\epsilon_{\alpha}$ is the energy density of the cold, beta-equilibrated barotropic EOS $\alpha$. Note that this proxy treats the NS matter as PNM at saturation, neglecting the proton and leptonic contributions \cite{Lattimer_2016}. 
We assume the proxy to be normally constrained, with $S_0=32$ MeV and $\sigma_0=2$ MeV \cite{Tsang_2012,Li_2019,Miller_2021};
our likelihood factor for a single EOS realization $\scY_{\alpha}$ is therefore 
\begin{equation}\label{eq:symenergy}
	\scS(\scY_{\alpha}) = \scL(S_0 | S_{\alpha}^{\mathrm{proxy}}(n_{sat})) = \scN(S_{\alpha}^{\mathrm{proxy}}(n_{sat}) | S_0, \sigma_0).
\end{equation} 
\noindent We obtain $S_{\alpha}^{\mathrm{proxy}}(n_{sat})$ by interpolating the rest-mass and energy densities extracted from our TOV solution. Kedia et al. \cite{atul2025} noted that this constraint favors smaller radii for any given $M_{max}$.


\section{Joint Inference Pipeline}\label{sec:hyperpipe}

To generate the posterior distribution from $\scL_{\rm net}$ and $p(\scY)$ as written in Eq. (\ref{eq:lnet}), we employ the \consf{HyperPipe} algorithm, a generalized version of RIFT originally introduced by Kedia et al. \cite{atul2025}. This pipeline extends the base RIFT package \cite{rift2018,Wysocki_2019b,expanding_rift2023,narrowing_rift2025} to include a general-purpose, physics-neutral implementation of the RIFT algorithm that can interface with generic codes, enabling it to handle non-GW model parameters $\scY$ and arbitrary, externally-supplied marginal evidence factors that appear in multi-event and multi-messenger inference, such as those in Eq. (\ref{eq:lnet}). The algorithm for \consf{HyperPipe} otherwise closely follows the original RIFT implementation and reuses many of the same components. 

\begin{figure}[t!]
	\centering
	\includegraphics[width=.48\textwidth]{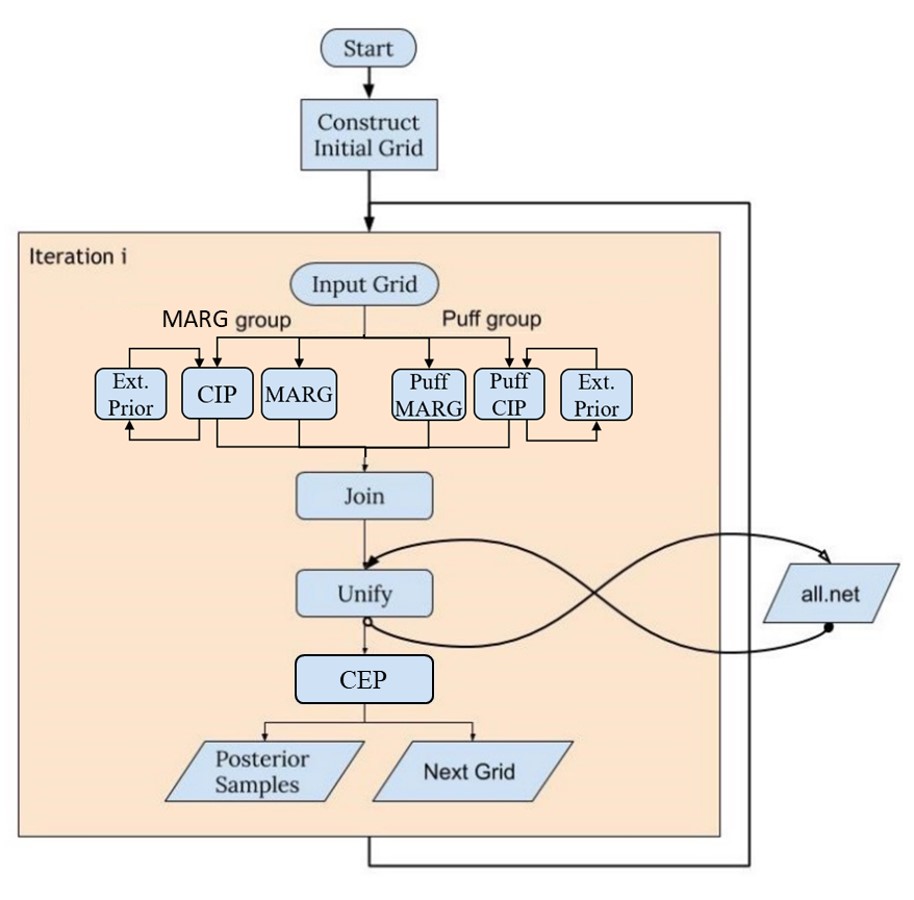}
	\caption{Flowchart for the iterative workflow of \consf{HyperPipe}, adapted from that for RIFT in \cite{expanding_rift2023}. The MARG and PUFF groups can contain multiple generic marginalization codes, such as CIP, that are simultaneously run hundreds of times in parallel to evaluate a selected number of samples from the input grid. We extend this stage of the algorithm by enabling the interaction between CIP and an external prior process that influences CIP's likelihood calculation. \label{fig:flowcharts}}
\end{figure}

In general, the algorithm adaptively explores the space of selected hyperparameters $\scY$ to resolve distinct features in the posterior distribution. We provide an initial grid of these parameters $\scY_{0}$ containing estimates within a subregion of the 7-dimensional space of $\scY = \{(\gamma_0, \gamma_1, \gamma_2, \gamma_3, \mu_1, \mu_2, \sigma)\}$ for its initial (index 0) iteration; these grid points are the values described in Sec. \ref{sec:methods}. The primary stages of each iteration are illustrated in the flowchart in Fig. \ref{fig:flowcharts}. We only focus on our modifications to these stages here; each stage is outlined in greater detail in \cite{thesis}.

Given an input grid of proposed hyperparameter sets $\scY_{\alpha}$ and the appropriate data, \consf{HyperPipe} first computes the marginal likelihood factors in Eq. (\ref{eq:lnet}) using the user-supplied modules. For this work, we employ three such modules, one for Eq. (\ref{eq:dk}), one for Eqns. (\ref{eq:nicer}) and (\ref{eq:symenergy}), and one for Eq. (\ref{eq:cip}). For this last factor, we repurpose the CIP module from RIFT mentioned in Sec. \ref{sec:gw}. 

We here extend \consf{HyperPipe} from its form in \cite{atul2025} by modifying the CIP module to interface with an external prior script that computes $p(x)$ for Eq. (\ref{eq:cip}) using a user-specified model. The interface is generalized such that the implementation will be the same regardless of the $p(x)$ model, providing broad flexibility. As our model is simply a normal distribution, we evaluate $p(x)$ for a given pair of masses $x=(m_1,m_2)$ as: 
\begin{equation}\label{eq:extp}
p(x) = \frac{1}{N_{\alpha}}\scN(x | \mu_{\alpha}, \sigma_{\alpha})
\end{equation}
\noindent where $N_{\alpha}$ is a normalization constant, defined as the multivariate CDF of $\scN$ within the valid region of our population model (see Fig. \ref{fig:popgrid}). The external prior module is depicted in Fig. \ref{fig:flowcharts} with a feedback loop connecting it to CIP alone.

In addition to the prior factor $p(x)$, this external module can provide CIP with a custom EOS solver and other likelihood factors, reducing reliance upon the current (limited) hardcoded functionality. To demonstrate this point, we also compute the pulsar maximum mass factor $\scC_{\alpha}$ (Eq. (\ref{eq:cdf})) in our external module, using EOS objects created via \consf{lalsuite} \cite{lalsuite,lal_paper}. Our code therefore supplies the product $\scC(\scY)\,p(x)$, such that: 
\begin{align}
	\scL_{\rm CIP} &= \int \scL_{marg}(X(x,\scY))\,f_{external}(x,\scY)\,dx \nonumber\\
	&= \int \scL_{marg}(X(x,\scY)) \left(p(x)\scC(\scY)\right)\,dx \nonumber\\
	&= \scC(\scY)\int \scL_{marg}(X(x,\scY))p(x)\,dx
\end{align}
\noindent Although this point is strictly technical, we highlight it to show the efficiency afforded by an external module, which reduces computing costs compared to running an entirely separate marginalization module for Eq. (\ref{eq:cdf}). 


In addition to this external prior functionality, which is mainly limited to working with the CIP module since it would be largely superfluous for other non-production codes, we implement two further technical modifications with broader impacts on the overall computational efficiency of \consf{HyperPipe}. Firstly, we implement the hypercube rotation described in Sec. \ref{sec:eos} into the posterior construction stage, denoted CEP, which performs a Monte Carlo integration to estimate the hyperparameter posteriors, similarly to Eq. (\ref{eq:riftpost}):
\begin{equation}\label{eq:cep}
p_{\rm post} = \frac{\scL(\scY)p(\scY)}{\int d\scY\, \scL(\scY)p(\scY)}
\end{equation}
\noindent In the original \consf{HyperPipe} implementation, the integration was performed over the $\gamma_k$ coordinates (and population hyperparameters) directly. Now, we first convert the $\gamma_k$ parameters to the better-aligned $r'_i$ coordinates, modifying \consf{HyperPipe} to access externally-provided coordinate transformation codes, specifically the transformation from \cite{dan2020}. Then, the CEP stage integrates over the $r'_i$ coordinates to more effectively sample posteriors for those hyperparameters, which are then converted back to $\gamma_k$ coordinates. 



We implement this same functionality in the final stage of \hyperpipe{}, denoted PUFF, which dithers the posterior grid to create a ``puffed'' grid that further explores the hyperparameter space. Specifically, we transform the puffed points into the rotated coordinate space to apply our second modification to \consf{HyperPipe}, which alters how the PUFF stage handles puffed points that fall outside the specified parameter ranges. Whereas the original version of \consf{HyperPipe} (and, indeed, RIFT) simply discarded these points, often substantially reducing the size of the puffed grid, we now \textit{reflect} them at the (buffered) $r'_i$ parameter boundaries, placing them back inside the hypercube $\scC'$ as a unique sampling point. As a result, we retain nearly all proposed puffed samples (culling only those for which $\mu_2 > \mu_1$), increasing the sampling density within $\scC'$ and enhancing the resolution of our posteriors. 

Note that, for our specific case, we only dither six of our seven total hyperparameters: we exclude the population width $\sigma$. Since it is poorly constrained by our observational data (see Sec. \ref{sec:validation}), all positive values of $\sigma$ remain physically valid, and puffed samples for this parameter would therefore broaden the widths of our Gaussian population model over multiple iterations, driving it towards uniform distributions in $\mu_1$, $\mu_2$ within our truncated ranges ($1\Msun \le \mu_1,\mu_2 \le 3 \Msun$).


Collectively, applying these rotation and reflection changes improves the overall efficiency of \hyperpipe{}, with the fraction of physical parameter sets in the first iteration jumping from $\sim 1$\% to $\sim 80$\% and the same fraction remaining at $ > 95\%$ in subsequent iterations. We further describe this behavior and how it alters the posteriors in Appendix \ref{sec:app}.





\section{Results \& Discussion}
\label{sec:results}

We first perform several tests to verify that the component modules behave as expected before running our modified pipeline in the full 7-dimensional configuration. 

\subsection{Validation Tests}\label{sec:validation}

\begin{figure}
	\centering
	(a)
	\includegraphics[width=.36\textwidth]{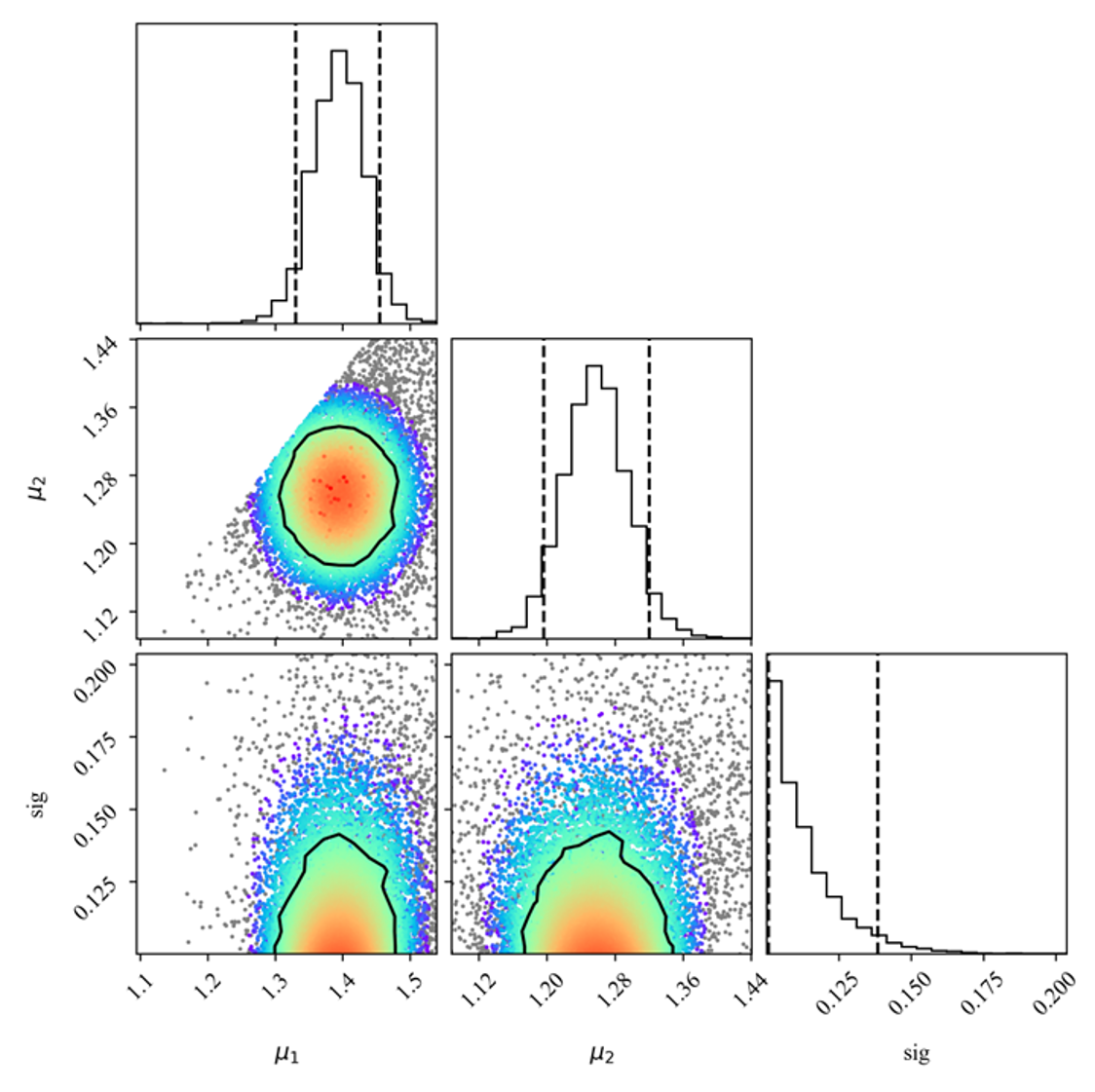} \\
	(b)
	\includegraphics[width=.36\textwidth]{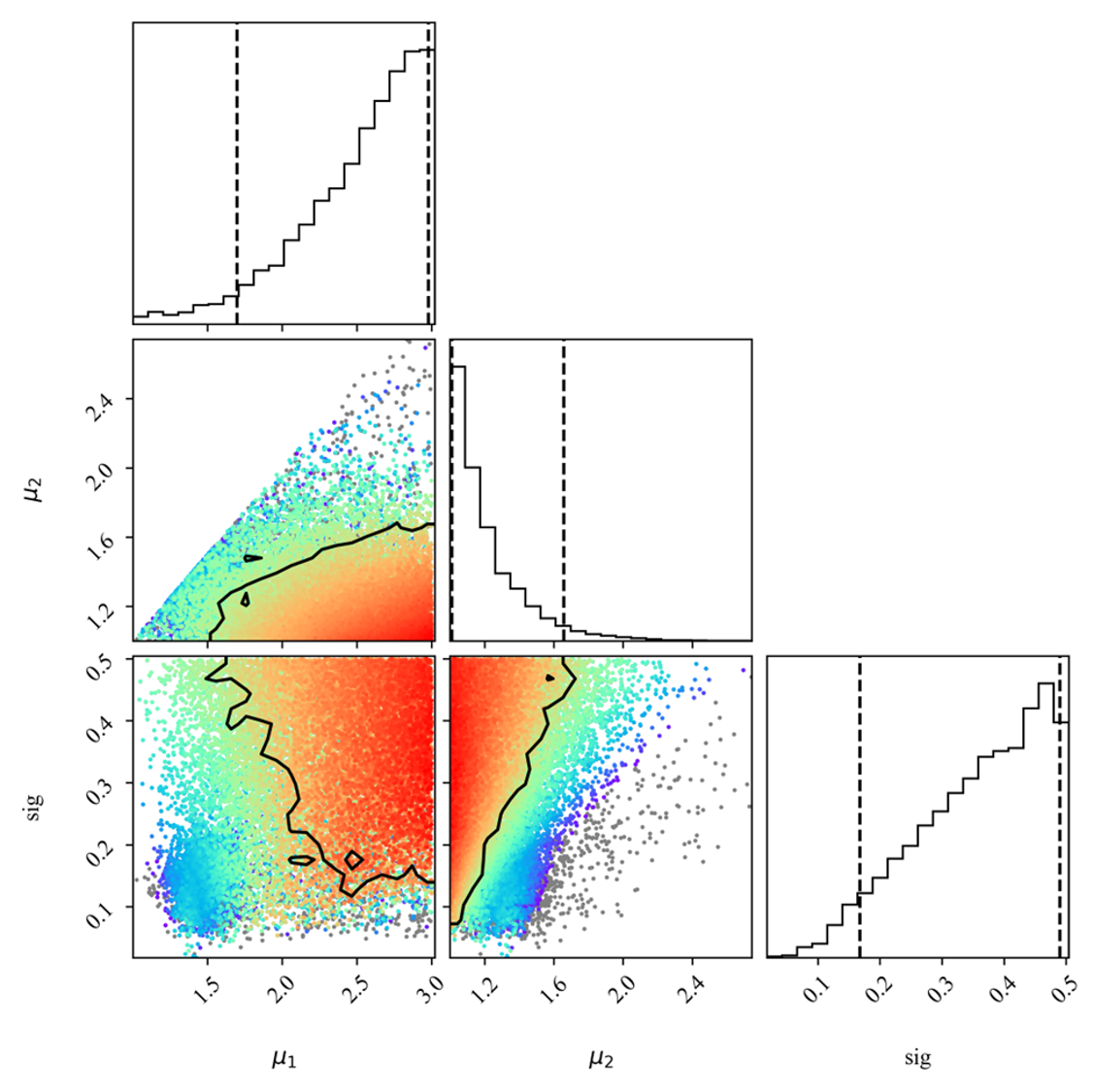} \\
	(c)
	\includegraphics[width=.36\textwidth]{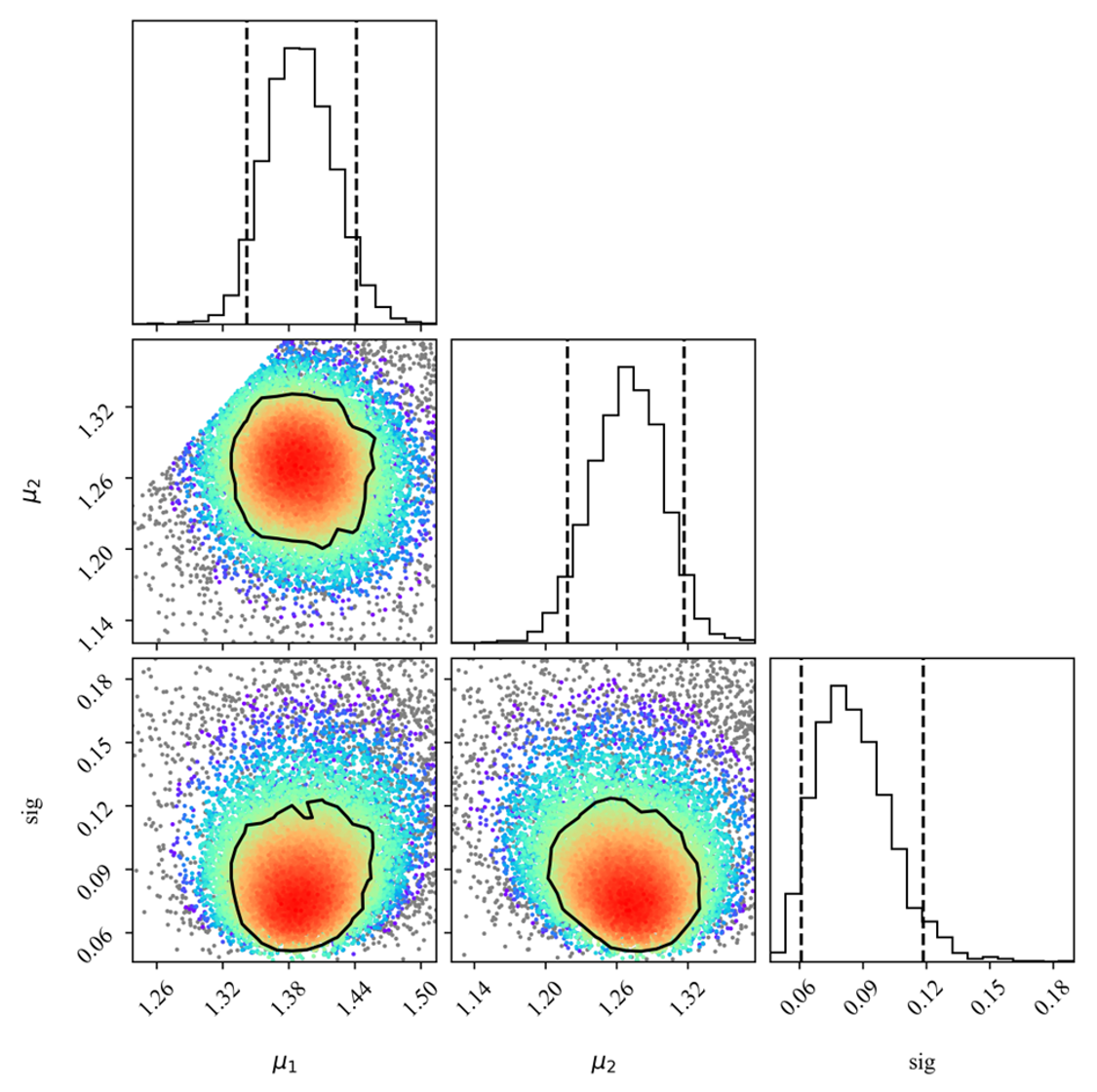}
	\caption{Corner plot posteriors for three population parameters $\mu_1$, $\mu_2$, and $\sigma$, generated over 3 \consf{HyperPipe} iterations using (\textit{a}) only galactic DNS mass constraints; (\textit{b}) only the GW170817 observation, interpreted with a flexible population but fixed EOS parameters; and (\textit{c}) both the DNS observations and GW170817 with fixed EOS and flexible masses. Under just the GW170817 constraint, the mass posteriors trend toward extremal mass ratio $q = \mu_2/\mu_1 \ll 1$, but they become well-constrained when the DNS observations are incorporated. \label{fig:mplt}}
\end{figure}

We look first at the behavior of the three population hyperparameters. Figure \ref{fig:mplt} shows three corner plot diagrams, each of which plots the 90\% credible interval of the converged posterior in black on top of the accumulated marginal likelihoods computed for all iterations in our analysis. A 1-D histogram for each parameter accompanies the scatterplots, with the dashed lines indicating the 5\% and 95\% quantiles for each parameter. 

The top panel of Fig. \ref{fig:mplt} shows the posterior obtained with only the DNS data likelihoods, calculated via Eq. (\ref{eq:dk}), included in the pipeline. As expected for an integral over the product of two Gaussians, the posterior also converges to a roughly normal distribution, centered on $(\mu_1,\mu_2) = (1.39 \pm 0.04 \Msun, 1.26 \pm 0.04 \Msun)$. For comparison, the sample means and widths for the 11 pairs of measured DNS masses listed in Table \ref{tbl:dns} are $(\mu_1,\mu_2) = (1.41 \pm 0.10 \Msun, 1.24 \pm 0.10 \Msun)$. Because the integral result drops off steeply the further $\mu_{pop}$ is from the region containing the real measurements, this factor constrains the sample population shown in Fig. \ref{fig:popgrid} tightly and drives the width $\sigma$ towards 0 (i.e., an unrealistic delta function population). Note that, because this code is completely independent of the EOS hyperparameters, it is essentially a population inference code only.

On the other hand, if we only use the CIP module with our external prior while holding the EOS parameters ``fixed'' - i.e., we apply the same fiducial set of EOS parameters to every set of population parameters for all iterations, neither fitting nor puffing the EOS values in \consf{HyperPipe} - the population posterior rails towards extreme mass ratio $q = \mu_2/\mu_1$, as shown in the center panel of Fig. \ref{fig:mplt}. The width $\sigma$ also rails towards large values, simply because broader distributions produce higher average likelihoods via Eq. (\ref{eq:extp}); note that we only integrate over the region $\sigma \in [0.01,0.499]$ in these tests, however. This behavior is expected, as there are no constraints placed on the mass values in this configuration, and the coordinate system we use in CIP favors extremal $q$.

Combining these two tests to incorporate both DNS and our prior via one fiducial GW170817 data point, as seen in the bottom panel of Fig. \ref{fig:mplt}, reproduces well-constrained posteriors for the population parameters, with somewhat broader distributions on the masses compared to the DNS-only test. The width $\sigma$ is now restrained by the competing effects of the two integrations, with a more Gaussian-shaped posterior, though it remains only weakly informed by the DNS samples. From these tests, we conclude that the population inference side of our pipeline is functional. 

\begin{figure}
	\centering
	\includegraphics[width=.47\textwidth]{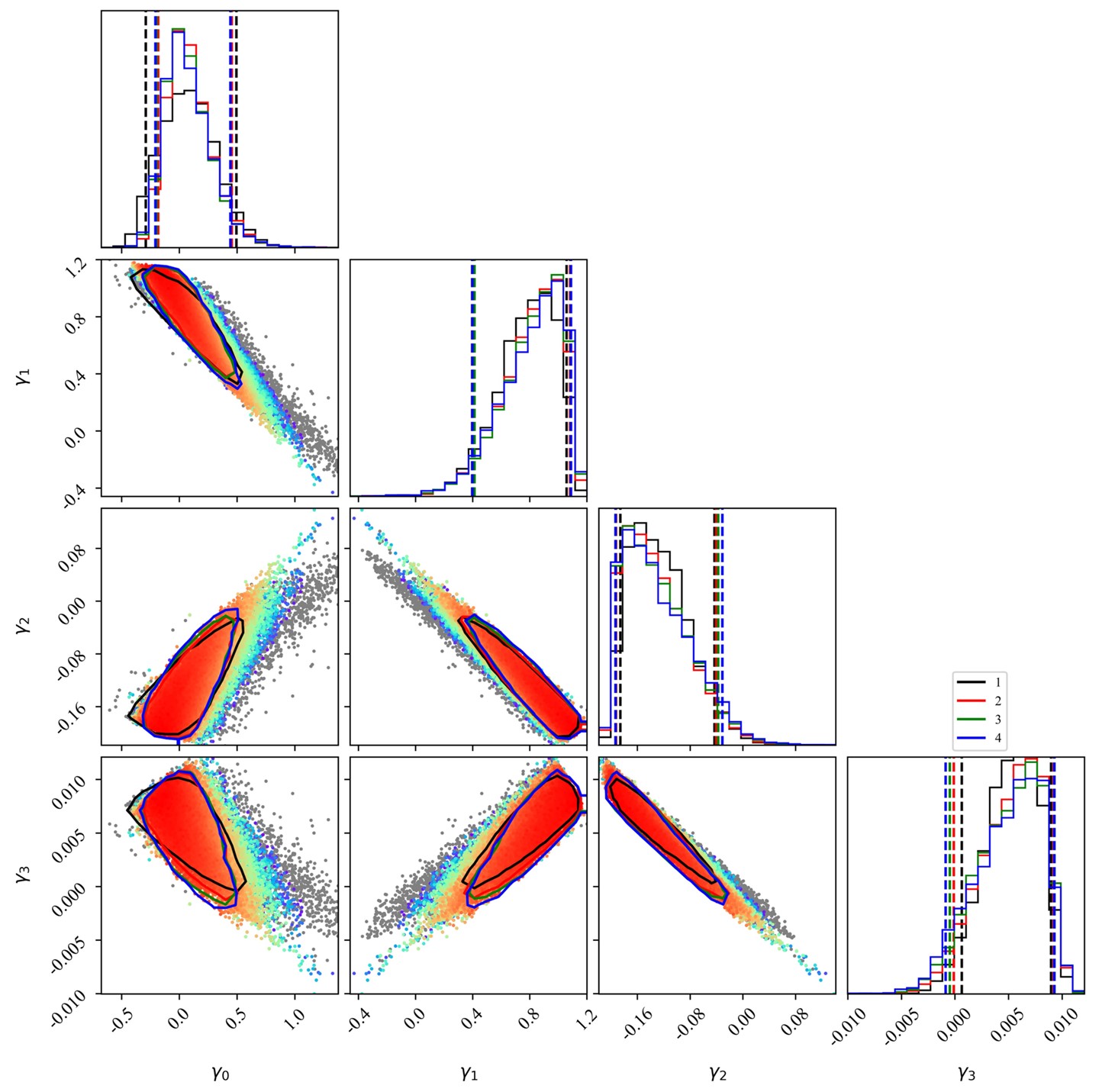}	
	\caption{Corner plot posteriors for the four $\Gamma$-spectral hyperparameters obtained over 4 iterations of the improved \consf{HyperPipe} algorithm, generated using data for GW170817 and NICER objects, alongside pulsar maximum mass and symmetry energy factors. Cf. Figure D.1 of \cite{atul2025}: the new posteriors are much better defined than those found by the original \consf{HyperPipe} algorithm, though they now rail towards one corner of the bounded physical space. \label{fig:CIPNCRcomp}} 
\end{figure}

On the other hand, we can also consider EOS inference on its own and compare our updated results to those of \cite{atul2025}, which utilized nearly identical data constraints, only omitting the NICER data for PSR-J0740, PSR-J0437, and the HESS J1731 source (\cite{Doroshenko_2022}). As pointed out in \cite{atul2025}, the HESS source only affects the posterior incrementally, consistent with the effect of the GW170817 measurement, so its inclusion does not strongly affect our comparison. 

Over 4 \consf{HyperPipe} iterations, we obtain the posterior samples and 90\% credible intervals shown in Fig. \ref{fig:CIPNCRcomp}, which quickly stabilize around a thoroughly-sampled region determined by the bounds of the rotated coordinate system (with 10\% buffer) detailed in Sec. \ref{sec:eos}. By contrast, the original \consf{HyperPipe} algorithm took 7 iterations to obtain sparsely-sampled, less coherent distributions for each of the hyperparameters: see Figure D.1 of \cite{atul2025}. Our rotated coordinates thus enable quicker recovery of stable, well-defined posteriors than the standard Cartesian $\gamma_k$ coordinates when performing EOS inference. 

Significantly, however, our posteriors for each hyperparameter appear to rail towards one corner of the hypercube in which every physical EOS lives. This behavior may be an outcome of using the non-causal $\Gamma$-spectral EOS parametrization, which allows the likelihood to peak in a non-physical region of the hyperparameter space. We address this behavior further below. 

\subsection{Joint Inference Analyses}\label{sec:result}

\begin{figure*}[t!]
	\centering
	\includegraphics[width=.98\textwidth]{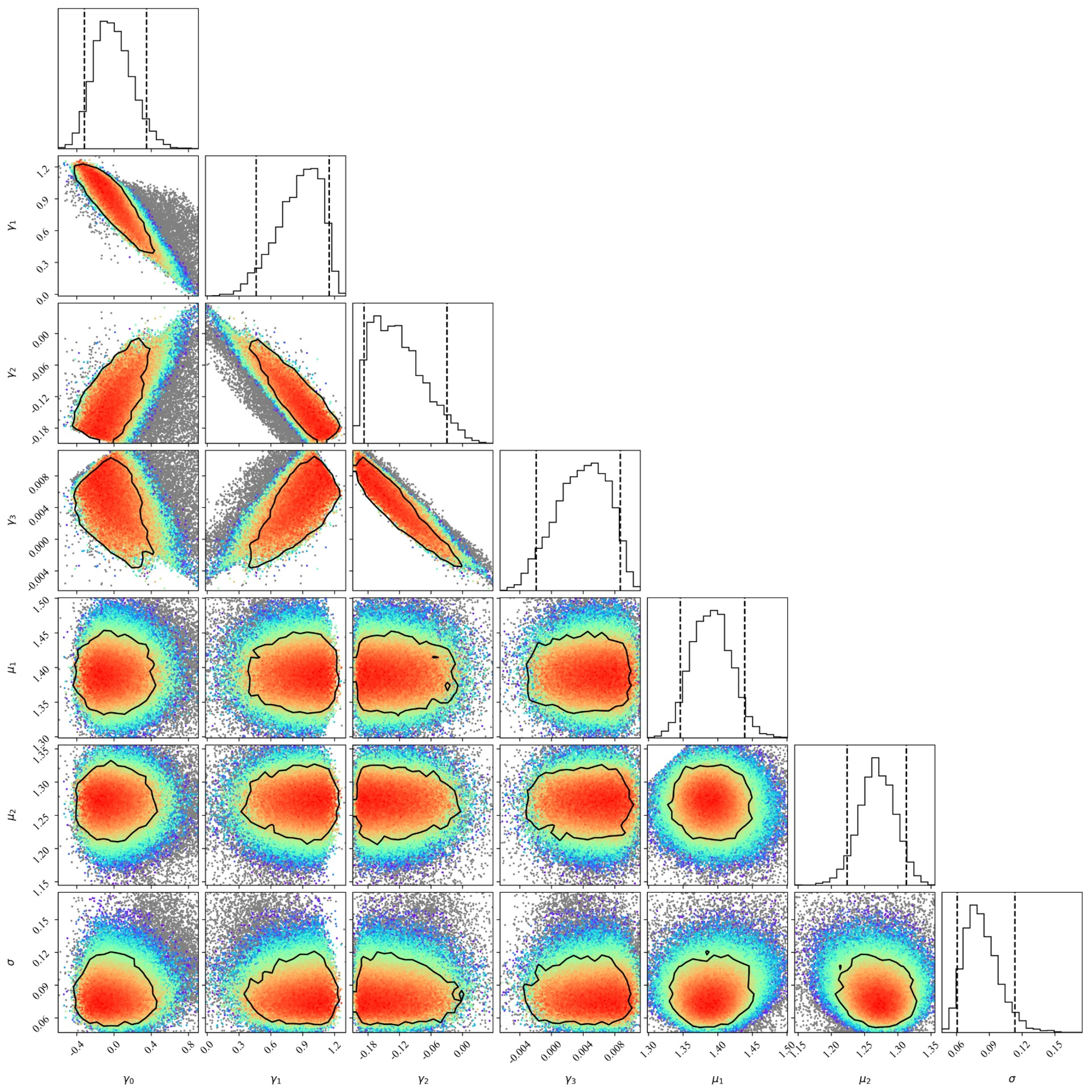}
	\caption{Corner plot posteriors for seven hyperparameters, generated over 6 \consf{HyperPipe} iterations with prior bounds buffered by 10\% and using constraint data from DNS masses, GW170817 with Gaussian population prior, high-mass and NICER pulsars and the HESS J1731 source, and the symmetry energy. The mass posteriors are constrained to Gaussians in line with the DNS mass weighting, while the spectral EOS parameters converge towards skewed distributions due to significant railing. \label{fig:all}}
\end{figure*}

\begin{figure*}[t!]
	\centering
	\includegraphics[width=.98\textwidth]{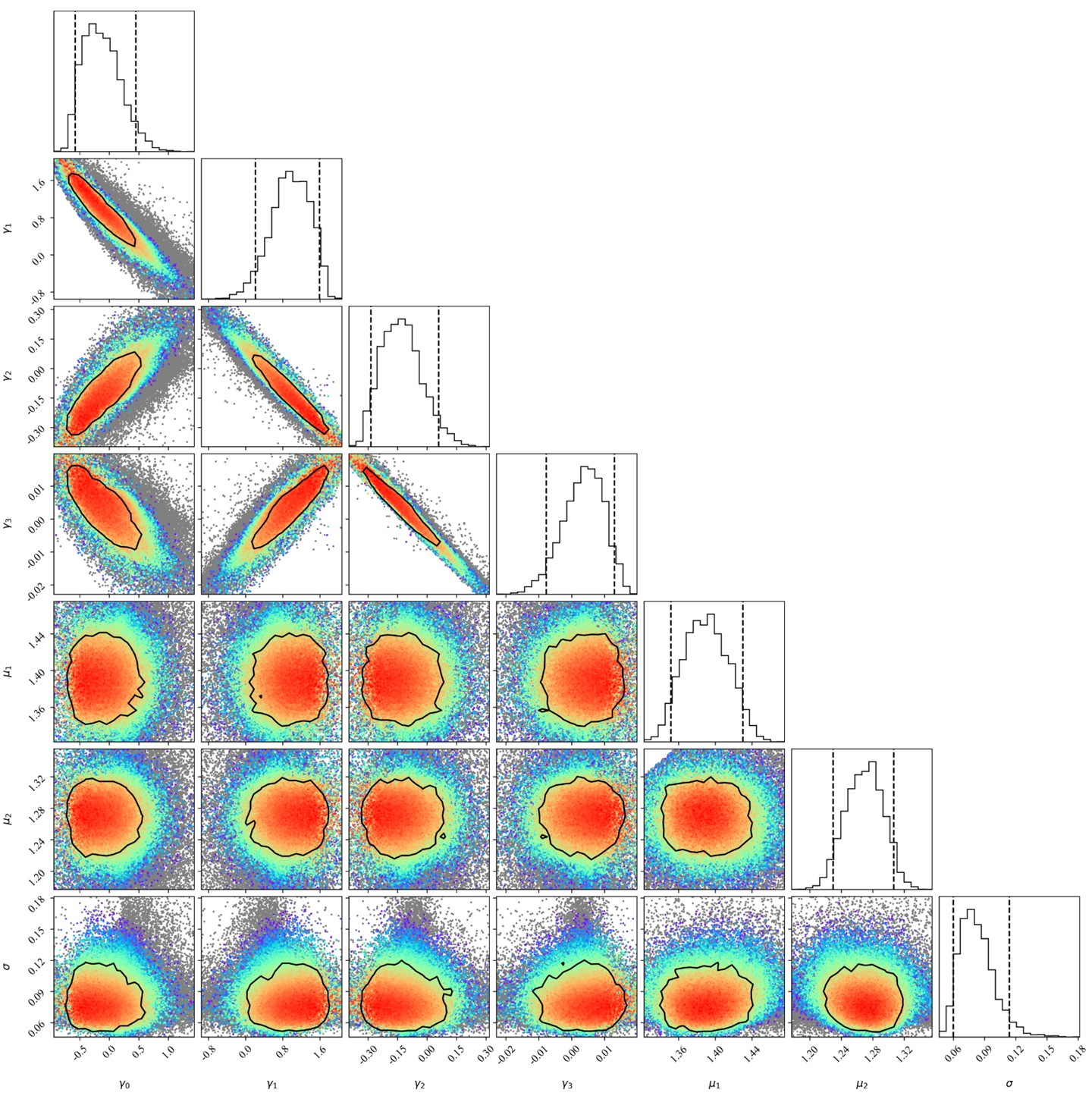}
	\caption{Corner plot posteriors for seven hyperparameters, generated over 15 \consf{HyperPipe} iterations, bounded by rotated coordinate prior bounds expanded by 400\%, and using the same data factors as in Fig. \ref{fig:all}. 
	The mass posteriors remain essentially the same as in Fig. \ref{fig:all}, while the EOS parameters are still skewed but no longer appear to be railing. \label{fig:all400}}
\end{figure*}

We turn now to performing full 7-dimensional joint inference of the neutron star EOS and BNS population hyperparameters $\scY = (\gamma_0,\gamma_1,\gamma_2,\gamma_3,\mu_1,\mu_2,\sigma)$, with all five constraining factors in Eq. (\ref{eq:lnet}) and all data sources incorporated into the analysis. Integrating over the parameters yields the 90\% credible posteriors shown in the corner plots in Figs. \ref{fig:all} and \ref{fig:all400} for two analyses, which apply a 10\% and a 400\% buffer factor to the hyperparameter ranges and which we hereafter denote as the $\scC_{10}'$ and $\scC_{400}'$ analyses, respectively.

\newcommand{\Cten}{\scC_{10}'}
\newcommand{\Cfour}{\scC_{400}'}

For the former analysis, the posteriors converged within 6 iterations of the \consf{HyperPipe} algorithm. The population hyperparameters (in the lower right section of the plot) converge as they did in the independent population inference test due to the strong weighting provided by the DNS population factor Eq. (\ref{eq:dk}), while the EOS hyperparameters (in the top half of the plot) are shifted somewhat away from the corner of the physical hypercube, unlike in the EOS-only test of Fig. \ref{fig:CIPNCRcomp}, but still continue to rail towards the hypercube edges. This shift demonstates the effect that inferring the population simultaneously with the equation of state has on the EOS hyperparameters, as indicated by \cite{dan2020}. We note that some sampled points have escaped our imposed physical hypercube bounds; this is a known artifact of the adaptive volume sampler \cite{AVsampler2023} we use. 

\begin{figure}[t!]
	\centering
	\includegraphics[width=.46\textwidth]{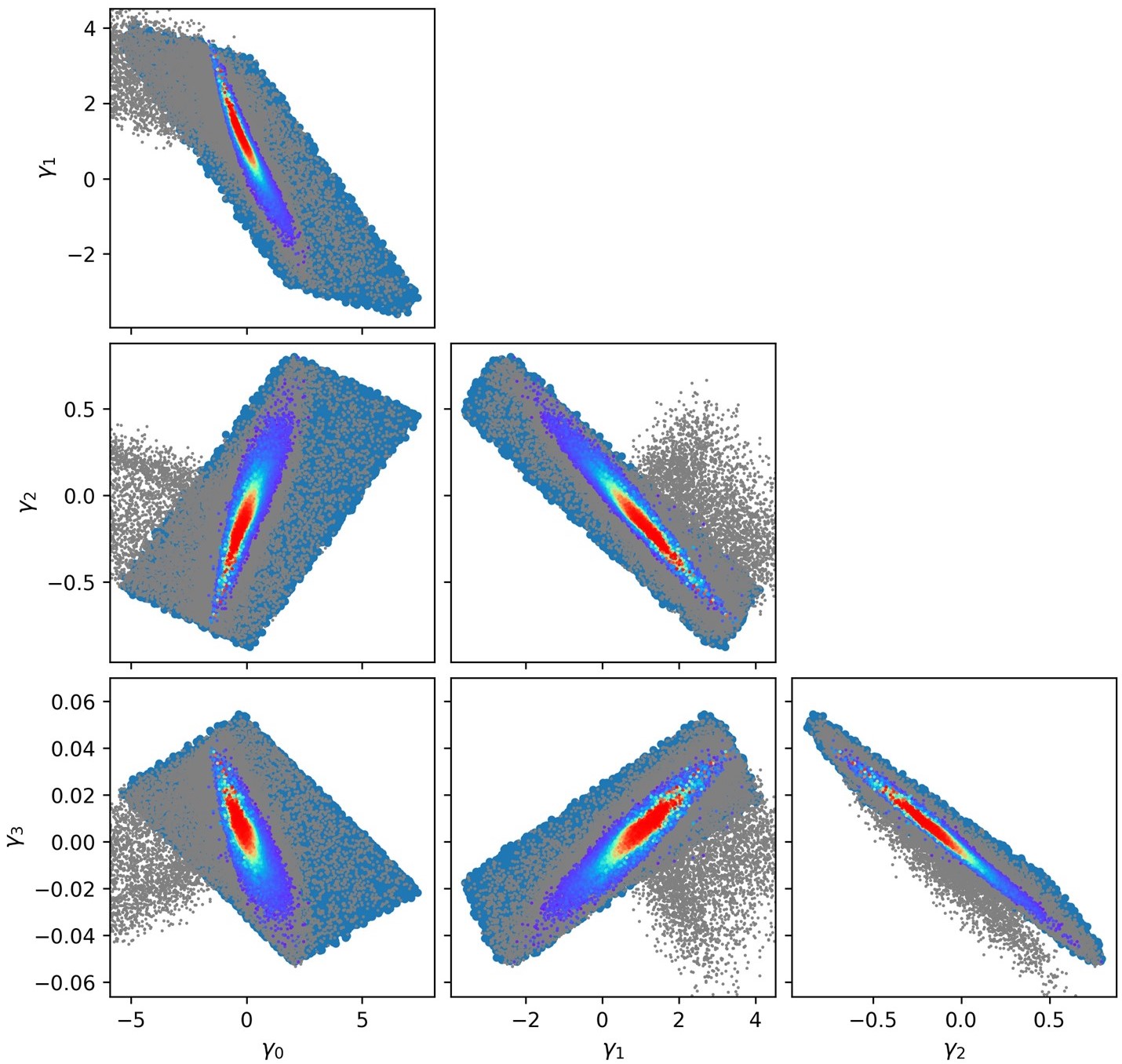}
	\caption{Corner plot of EOS hyperparameter samples for the $\Cfour$ analysis, overlaid on points showing the 400\%-buffered hypercube. Colored points represent physical EOS realizations, while grey points are unphysical samples. The physical region is fully enclosed by the hypercube bounds and do not appear to be railing against the edges. \label{fig:box}}
\end{figure}

\begin{figure}[ht!]
	\centering
	\includegraphics[width=.44\textwidth]{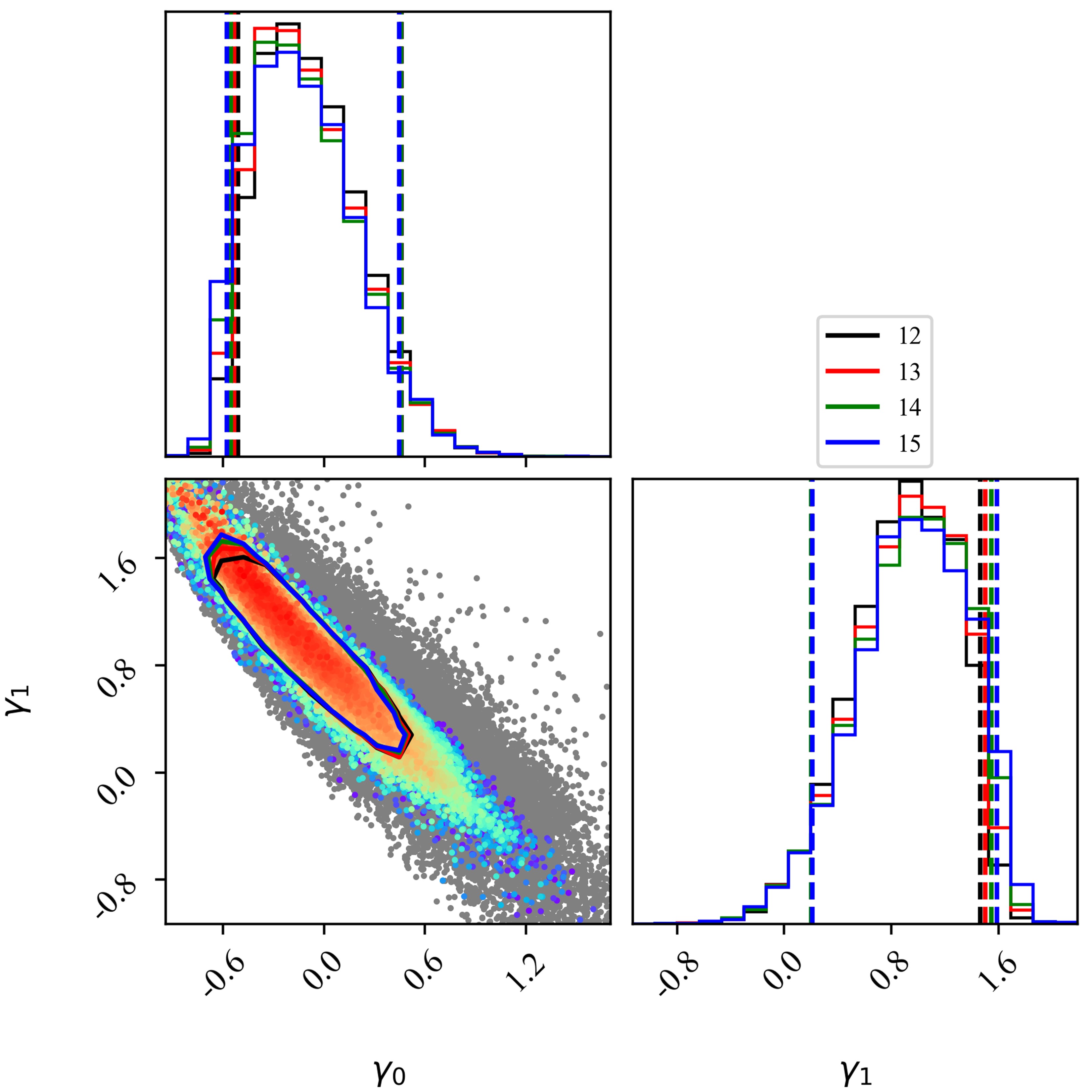}
	\caption{Corner plot posteriors for two EOS hyperparameters over the final 4 iterations of the $\Cfour$ analysis. 
	While mostly stable, the posterior grows slightly with each iteration toward the upper-left corner, where the sample likelihoods stop scaling continuously. \label{fig:stability}}
\end{figure}

On the other hand, the $\Cfour$ analysis (Fig. \ref{fig:all400}) ran for 15 iterations. While the population parameter posteriors remain largely identical to the $\Cten$ analysis - the buffer factor does not affect those hyperparameters' bounds - the EOS parameter posteriors no longer appear to be railing, and instead 
lie well inside the bounds of the 400\%-buffered rotated hypercube $\scC'$, as demonstrated by Fig. \ref{fig:box}, where the samples tested in the $\Cfour$ analysis are overlayed on many draws that illustrate the shape and size of $\scC'$. 
Whether the hyperparameters would still rail, given many more iterations, is uncertain. It is possible that \consf{HyperPipe} did not explore the space quickly enough for the posteriors to reach the edge of the larger hypercube within 15 iterations: as shown in Fig. \ref{fig:stability}, the posteriors appear to have mostly stabilized after 12 iterations, expanding only slightly towards the upper left corner of the $\gamma_0-\gamma_1$ scatterplot with each additional iteration. However, the likelihoods of the samples in that upper-left region do not smoothly transition to smaller values. This ragged end may indicate the edge of the physical space, as our TOV solver enforces its inherent causality constraint, or it simply suggests this region was not sampled thoroughly enough during the analysis (despite the high number of samples already drawn there, as seen in Fig. \ref{fig:box}). For the remainder of this work, we take these posteriors to have indeed stabilized around the fully-sampled physical space of the spectral EOS model, as inferred using our data sources.

\begin{figure}[t!]
	\centering
	\includegraphics[width=.47\textwidth]{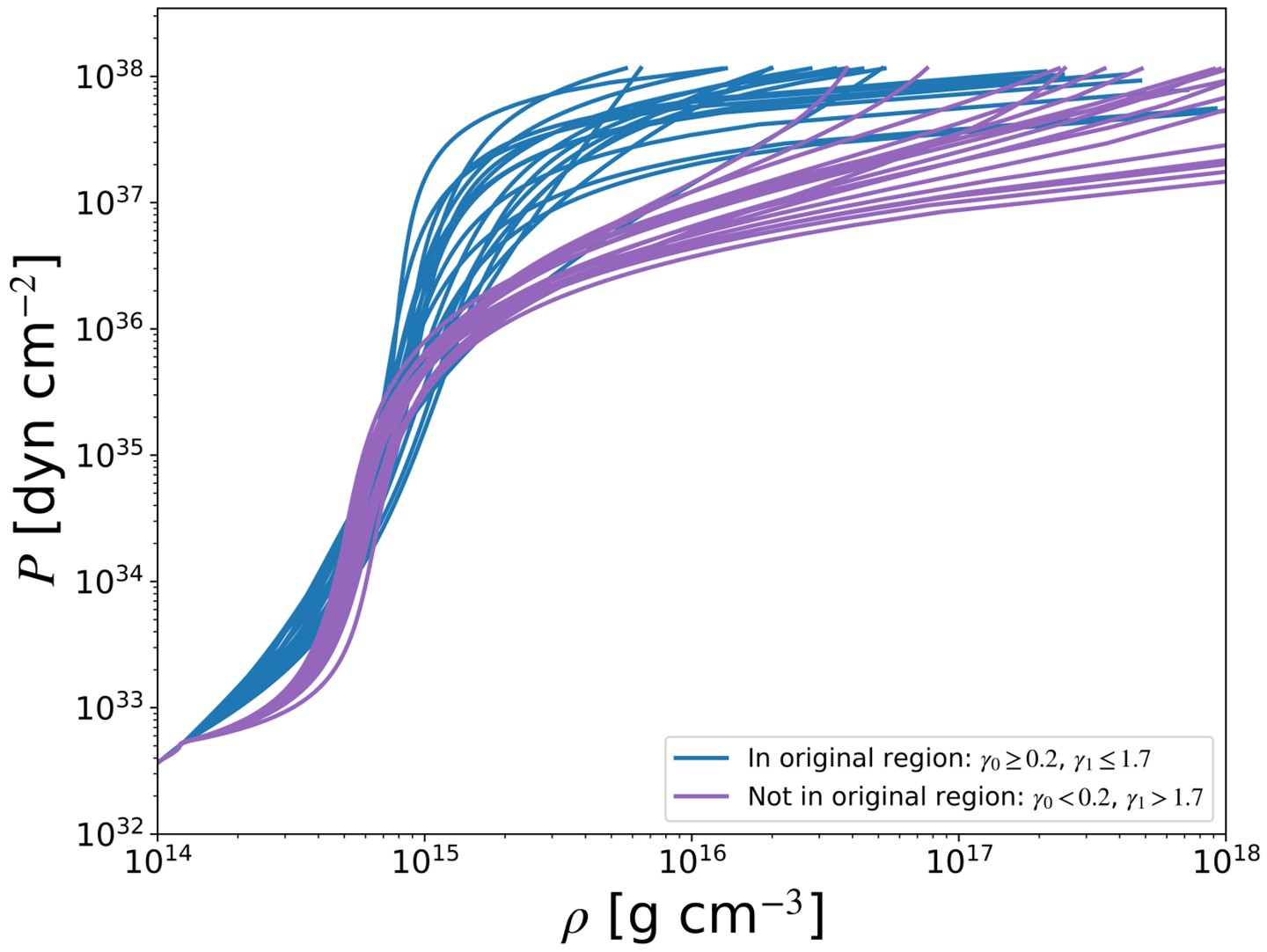}
	\caption{$P(\rho)$ curves for 40 sets of EOS parameters drawn from the final posterior of the $\Cfour$ analysis. \textit{Blue}: 20 sets with all $\gamma_k$ parameters within the original Cartesian region defined by \cite{Carney_2018}. \textit{Purple}: 20 sets lying outside that original region in two dimensions, with $\gamma_0 < 0.2$ and $\gamma_1 > 1.7$. \label{fig:oob}}
\end{figure}

Note that the original Cartesian bounds for the $\gamma_k$ hyperparameters, as proposed in \cite{Carney_2018} and given in the top row of Table \ref{tbl:bounds}, exclude physically-valid samples contained in these posteriors, and 
those EOS realizations 
produce markedly different EOS posterior curves. Figure \ref{fig:oob} plots EOS pressure-density curves for 40 realizations drawn from the hyperparameter posteriors of Fig. \ref{fig:all400}, 20 from within the standard Cartesian bounds (shown in blue) and 20 that are partially outside these bounds, having values for $\gamma_0 < 0.2$ and $\gamma_1 > 1.7$ (shown in purple).\footnote{Even with expanded bounds in rotated coordinates, the $\gamma_2$ and $\gamma_3$ samples mostly fit within the original Cartesian bounds; as a result, in our final posteriors, there were no EOS realizations with all four $\gamma_k$ coordinates outside the original bounds.} 
The latter curves exhibit a substantial change in sound speed at the matching density and lower pressures at high density. 

\begin{figure*}[t!]
	\centering
	\includegraphics[width=.49\textwidth]{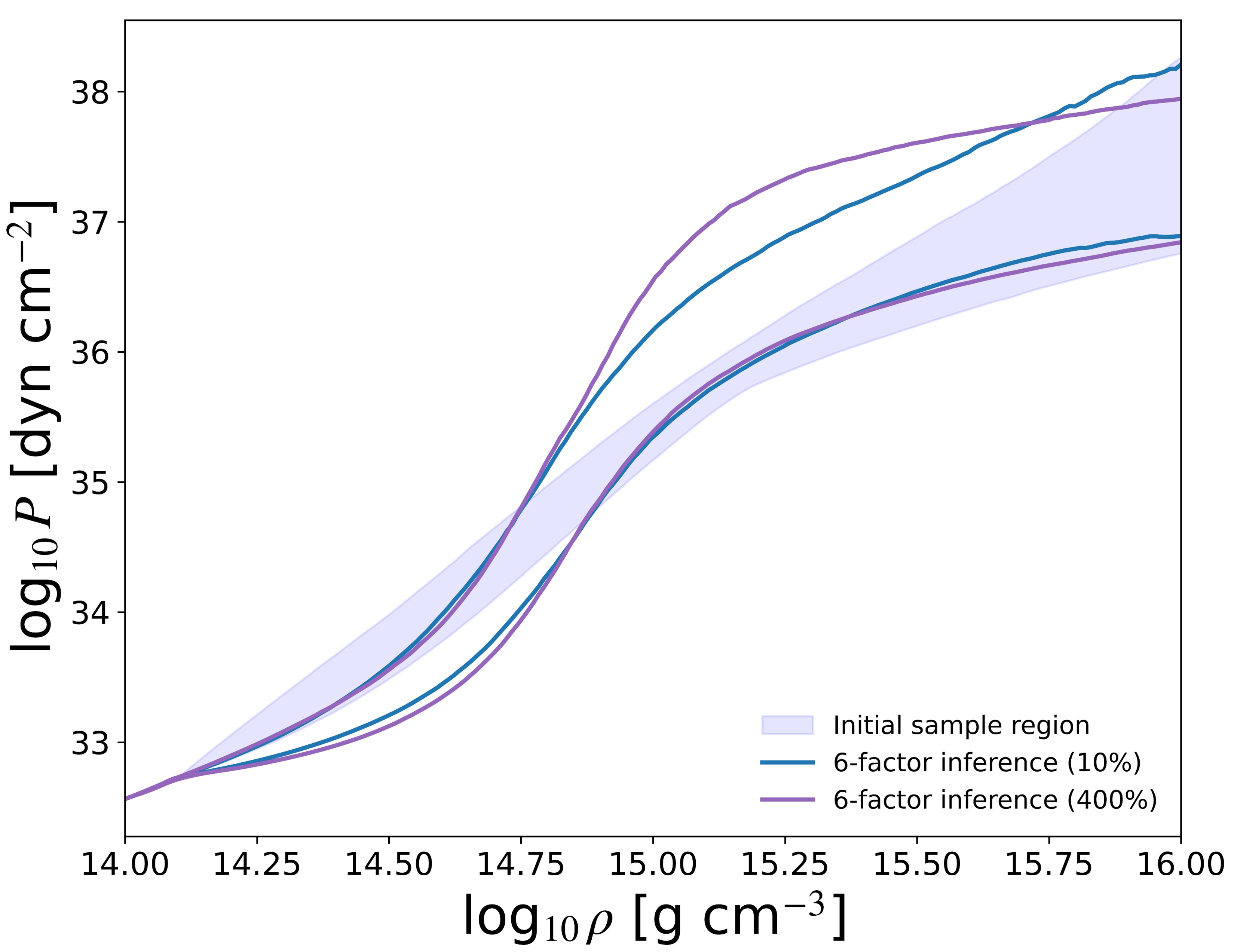}	
	\includegraphics[width=.49\textwidth]{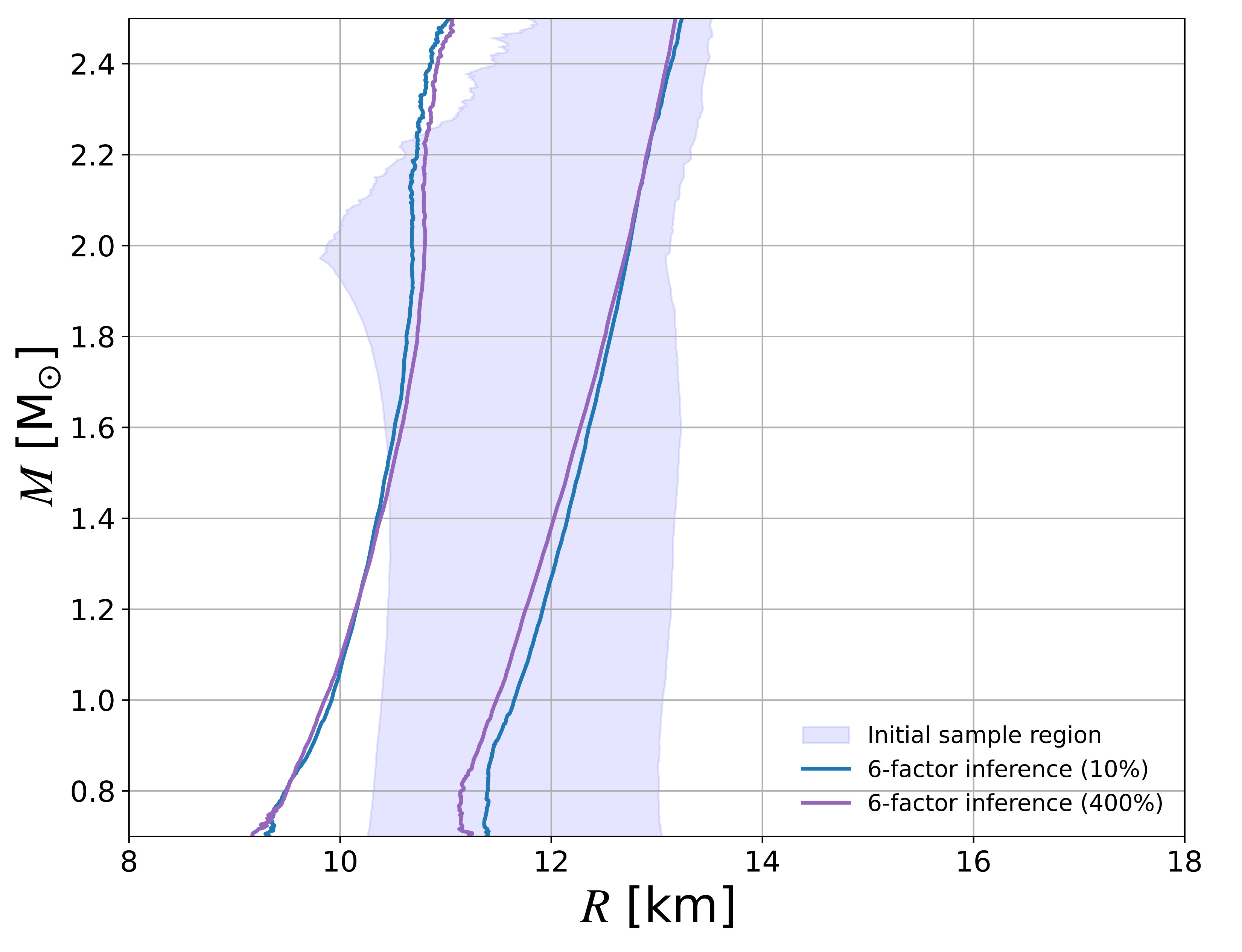}
	\caption{Joint 6-factor EOS constraint from DNS systems, galactic radio and x-ray pulsars, symmetry energy, and GW170817 using the 4-dimensional uniform-prior $\Gamma$-spectral EOS parametrization and a 3-dimensional truncated Gaussian DNS population prior. The blue and purple lines bound the 90\% confidence intervals for the 10\% and 400\% buffered EOS parameter ranges, respectively. We provide a comparison with our initial EOS samples drawn from a posterior for GW170817. Our posteriors indicate a sharper shift from low to high pressure as density increases and put tighter constraints on the possible radii at each mass. \label{fig:PDMR}}
\end{figure*}

The joint EOS constraints associated with both analyses $\Cten$ and $\Cfour$ are shown in Fig. \ref{fig:PDMR}, with the initial EOS samples from GW170817 (\cite{170817EOSdata}) indicated by the shaded region, for comparison. 
We refer to these analyses as ``6-factor inferences,'' distinguishing between \consf{HyperPipe}'s base GW inference module, which defaults to uniform mass priors, and our external prior mass model, since it is a central focus of this work. 
We find both posteriors to be roughly similar in shape in both the pressure-density and mass-radius plots. 
The key differences are inherited from the extra hyperparameter configurations afforded by the $\Cfour$ analysis and
illustrated in Fig. \ref{fig:oob}: the pressure at low density is marginally lower, the sound speed at high
density is generally smaller, and the pressure has a more pronounced change at densities around $\log_{10}\rho \sim 14.8$ g/cm$^3$. 
Meanwhile, the mass-radius posteriors indicate a more tightly constrained bound than the initial samples, which is expected since we incorporated both an independent mass prior and data from DNS sources that directly influenced the mass parameters. 
In particular, our analyses favor smaller radii ($R \lesssim$ 10-12 km) at low and subsolar masses. 

Given the similarity of our two analyses' results, for the remainder of this discussion we consider just the results of the $\Cten$ inference. 

\begin{figure*}[t!]
	\centering
	\includegraphics[width=.48\textwidth]{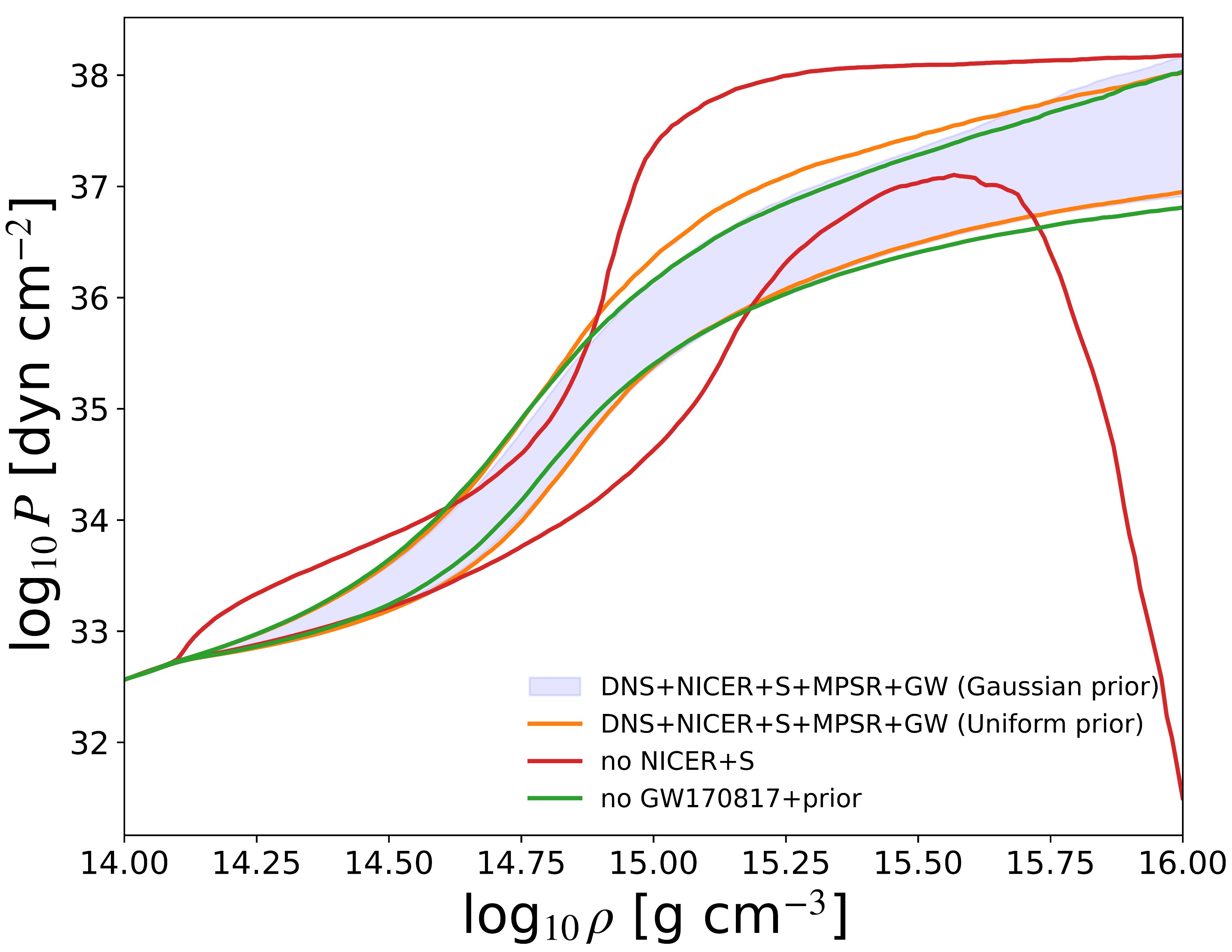}	
	\includegraphics[width=.48\textwidth]{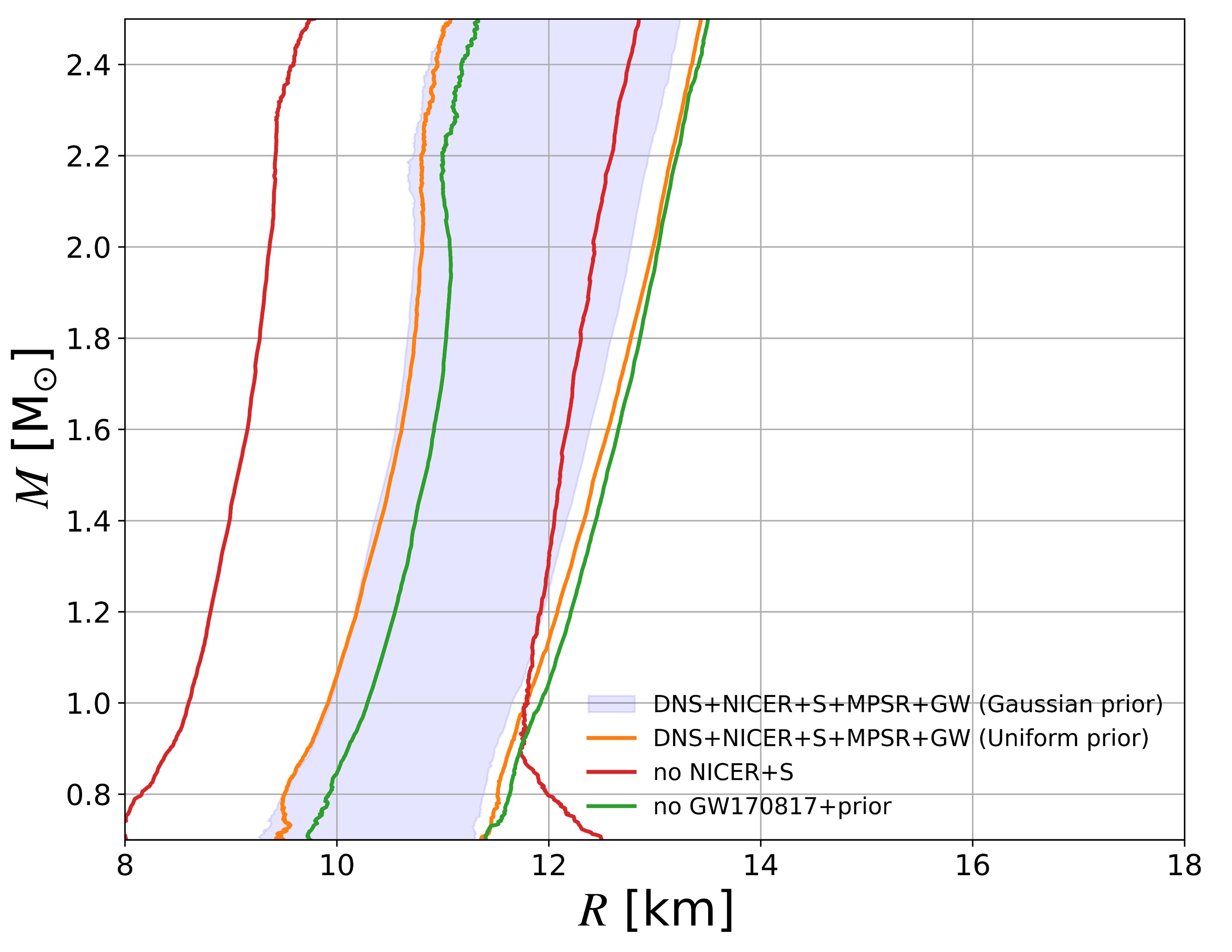}
	\caption{Joint EOS constraint at 90\% confidence intervals, with individual marginal likelihood factors of Eq. (\ref{eq:lnet}) removed: the external prior, the NICER data and symmetry energy, and the GW data with external prior. The 6-factor inference posterior from Fig. \ref{fig:PDMR} is plotted as the shaded region. Removing our Gaussian prior to use a uniform one produces minute changes in both plots, while removing the GW inference and prior altogether shifts the results slightly more. Excluding the NICER observations creates a much less tightly constrained distribution in both plots. \label{fig:pdmrcomp}}
\end{figure*}

It is worth comparing the result of this EOS inference with the results when certain factors are removed from Eq. (\ref{eq:lnet}), to observe their effects on the posterior distributions. 
In particular, as demonstrated by \cite{dan2020}, the choice of population model should have a significant impact on the inferred EOS. 
However, if we disable our external prior code, such that the GW inference module reverts to its default uniform mass priors, we find that the EOS inference is only shifted slightly by the prior, as seen in Fig. \ref{fig:pdmrcomp}, where the uniform-prior results are plotted in orange and the 6-factor results are shown as the shaded region. 
This outcome is not surprising, though, since our prior model informs the EOS solely through the gravitational wave likelihood factor Eq. (\ref{eq:cip}), which itself carries relatively little weight in our specific analysis because we have included only one GW measurement, GW170817. 
As shown by the green line in Fig. \ref{fig:pdmrcomp}, the EOS inferred without the GW170817 data (and therefore without the prior) only shifts the $M(R)$ and $P(\rho)$ posteriors incrementally. 
Thus, while this choice of prior has some effect on the population inference (see Fig. \ref{fig:mplt}), its impact on the EOS is limited by the number of gravitational wave detections involved in the inference. 
A more complex population model, as well as additional GW events, will induce more interesting behavior, but this test at least demonstrates that \consf{HyperPipe} will correctly handle those cases.
	
In contrast to the GW data, the NICER data imposes much stronger constraints on the EOS, as evidenced by comparing the green 90\% confidence intervals in Fig. \ref{fig:pdmrcomp}, which exclude the GW data, with the red bounds in the figure, which exclude the NICER data and symmetry energy. 
Without these factors, the EOS is clearly much more poorly constrained than in our 6-factor inference, indicating that the GW170817 data with external prior and the NS maximum mass constraint (Eq. (\ref{eq:cdf})) alone do not tightly constrain the EOS, particularly at high density. 
Also, the NICER data supports stiffer EOS models than the GW data, since its inclusion pulls the posterior's lower mass-radius bound to larger radii for each mass. 
Again, though, while the NICER data currently has the largest influence on the EOS, future GW measurements should shift this balance of power towards more equal weight between these data sources.

Finally, we can look quantitatively at our novel results for the population parameters, which have not been inferred in this manner before now. Previous studies have typically reported just one parameter $\mu_{dns}$, assuming a unimodal population distribution from which both members of a binary NS system are independently drawn, but this approach disregards the possibility that the individual members of the pair could have strongly correlated masses due to their astrophysical formation channels \cite{dan2020}. In particular, \cite{Alsing_2018} considered several Gaussian mixture models for the population and found that the $n = 1$ (i.e., unimodal) model was strongly disfavored compared to higher-component models, most notably that with $n=2$. In our analysis, we have employed a separate unimodal model on each member of an NS binary system, $\mu_1$ and $\mu_2$, treating them as independent hyperparameters and allowing \consf{HyperPipe} to determine the shape of the population for itself, based on the data we provided. For the 6-factor inference posteriors shown in Fig. \ref{fig:all}, our population hyperparameters have mean values $\mu_{1,pop} = 1.39 \pm 0.03 \Msun$, $\mu_{2,pop} = 1.27 \pm 0.03 \Msun$, and $\sigma_{pop} = 0.08 \pm 0.02 \Msun$. 

\begin{table*}
	\caption{Population means and dispersions for the NS mass distribution as reported by several studies. Note that we relabel the masses in the bimodal models such that $\mu_1 \ge \mu_2$, for consistency with our notation. \label{tbl:popres}}
	\begin{center}
		\begin{tabular}{l ccccc}
			\hline
			Work & Model & $\mu_1\,[\Msun]$ & $\mu_2\,[\Msun]$ & $\sigma_1\,[\Msun]$ &$\sigma_2\,[\Msun]$\\ 
			\hline & & & & & \vspace{-.25cm}\\
			\vspace{.1cm}This work & 2-D normal & $1.39^{+0.03}_{-0.03}$ & $1.27^{+0.03}_{-0.03}$ & $0.08^{+0.02}_{-0.02}$ & $0.08^{+0.02}_{-0.02}$ \\ 
			\vspace{.1cm}Anik et al. \cite{anik2025} & Skewed unimodal normal & $1.39^{+0.04}_{-0.07}$ & --- & $0.14^{+0.04}_{-0.01}$ & --- \\
			\"Ozel et al. \cite{Ozel_2012} & \makecell{Unimodal normal \\2-D normal} & \makecell{1.33 \\ 1.35} & \makecell{--- \\ 1.32} & \makecell{0.05 \\ 0.05} & \makecell{--- \\ 0.05} \\
			\vspace{.05cm}Kiziltan et al. \cite{Kiziltan_2013} & Skewed unimodal normal & 1.32 & --- & 0.225 & --- \\
			\vspace{.1cm} Alsing et al. \cite{Alsing_2018} & Truncated bimodal Gaussian & $1.80^{+0.15}_{-0.18}$ & $1.34^{+0.03}_{-0.02}$ & $0.21^{+0.18}_{-0.14}$ & $0.07^{+0.02}_{-0.02}$ \\ 
			\vspace{.1cm}Antoniadis et al. \cite{Antoniadis_2016} & Bimodal Gaussian & $1.807^{+0.081}_{-0.132}$ & $1.393^{+0.031}_{-0.029}$ & $0.177^{+0.115}_{-0.072}$ & $0.064^{+0.064}_{-0.025}$ \\
			\hline 
		\end{tabular}
	\end{center}
\end{table*}

Comparing these results with those of other population inference studies, listed in Table \ref{tbl:popres}, we find that our mass distribution is broadly consistent with the unimodal distribution previously reported to peak in the range of $\approx 1.32-1.39 \Msun$ \cite{Ozel_2012,Kiziltan_2013,anik2025}, from which both masses in a given binary are considered to be drawn independently. In particular, \cite{Ozel_2012} tested a bivariate population indentical to ours, with independent distributions for primary and companion masses, and found parameters $\mu_{1,pop} = 1.35 \Msun$, $\mu_{2,pop} = 1.32 \Msun$, and $\sigma_{pop} = 0.05 \Msun$. That work concluded that both masses were most likely drawn from the same narrow distribution; although our masses are slightly further apart, they too remain within one population width $\sigma_{pop}$ of each other, so our results do not definitively indicate a bimodal normal distribution for the population. Indeed, if we assume our mass posteriors for each hyperparameter are equally likely, we can average them to obtain a single distribution with peak $\mu = 1.33 \pm 0.08 \Msun$, in close agreement with the unimodal result of \cite{Ozel_2012}. 

We can also consider the more robust bimodal distributions determined by \cite{dan2020,Antoniadis_2016,Alsing_2018}; clearly our results show no indication of a secondary (or tertiary, if considering $\mu_{1,pop}$ and $\mu_{2,pop}$ as belonging to distinct distributions) higher-mass peak, since we assumed only a unimodal distribution for our prior model, but our $\mu_{1,pop}$ value agrees well with their lower-mass peaks, particularly that of \cite{Antoniadis_2016}. It is possible that incorporating more X-ray and NS-WD system measurements into the likelihood calculation and employing a bimodal normal prior in each of our hyperparameters could reveal such bimodality within each of our two-dimensional posteriors, and \consf{HyperPipe} is fully capable of handling such a complex model. However, such an analysis would no longer be inferring just the \textit{double} neutron star population but instead a universal NS mass distribution, and we have not accounted for the various selection effects that arise in that scenario \cite{anik2025}. 




\section{Conclusions \& Future Work}\label{sec:conclusion}

We have extended \consf{HyperPipe} to incorporate externally-evaluated prior
factors into RIFT's gravitational-wave marginalization and to support arbitrary
coordinate transformations when constructing and exploring hyperparameter
posteriors. We applied this framework to a seven-parameter joint inference of
the neutron star EOS and binary neutron star mass distribution, combining a
four-parameter $\Gamma$-spectral EOS decomposition with a truncated bivariate normal mass model. The analysis included constraints from galactic double neutron stars, GW170817, massive radio pulsars, NICER X-ray pulsars and the HESS J1731 source, and the nuclear symmetry energy.


For the population parameters, the joint analysis inferred $\mu_{1,\mathrm{pop}}=1.39\pm0.03\,\Msun$,
$\mu_{2,\mathrm{pop}}=1.27\pm0.03\,\Msun$, and
$\sigma_{\mathrm{pop}}=0.08\pm0.02\,\Msun$. 
These values are broadly consistent with previous unimodal estimates of the double neutron star mass distribution~\cite{anik2025,Ozel_2012,Kiziltan_2013,Ghosh_2025} and do not by themselves require distinct mass subpopulations. 
However, of the present data set, the mass-radius measurements of the NICER pulsars had the most significant constraining effect on the EOS posterior, while the population prior had only a minor effect since GW170817 was the sole gravitational-wave event coupling the mass distribution to the EOS.  
With additional binary neutron star merger detections, the population model should more strongly inform the joint EOS inference.
Regardless, this test establishes the machinery needed to incorporate more complex population models, including the mixture distributions considered in previous work~\cite{dan2020,Alsing_2018,Antoniadis_2016}, into the pipeline structure.

Additionally, the new mechanisms for coordinate rotation and reflection of exploratory samples allowed \consf{HyperPipe} to obtain densely sampled, stable EOS posteriors more quickly than the unmodified Cartesian workflow. 
Expanding the rotated EOS-parameter bounds from \cite{dan2020} also removed the visible EOS boundary-railing in our primary analysis while leaving the population posterior nearly unchanged, exposing physical EOS realizations outside the standard Cartesian bounds that have been in use since \cite{Carney_2018}.


Future \consf{HyperPipe} applications should incorporate the associated population selection effects into the mass inference, such as those for radio pulsar surveys, which Drummond et al. \cite{Drummond_2026} recently found to have negligible impact on the recovered mass distribution but significant effect on other binary parameters. In addition, other data factors could substantially influence the inference, such as high-density constraints from perturbative QCD analyses \cite{qcd} and additional (e.g., \cite{Mauviard_2026}), alternate (e.g., \cite{Riley_2019,Riley_2021,Salmi_2022}) or updated (e.g., \cite{Vinciguerra_2024,Dittmann_2024}) mass-radius estimates for MSPs from NICER data sets.
Finally, further tests could also examine the sensitivity of the EOS posterior to alternative interpretations of the low-mass HESS J1731 compact object~\cite{Hovarth_2023,J1731new,Pal_2025}; 
mass-radius estimates of this source have so far been noted to only have an incremental effect on the posterior, roughly consistent with that of gravitational wave events \cite{atul2025}.


\begin{acknowledgments}
	The authors are grateful for computational resources provided by the LIGO
	Laboratory and supported by National Science Foundation Grants PHY-0757058 and
	PHY-0823459. This material is based upon work supported by NSF's LIGO
	Laboratory, which is a major facility fully funded by the National Science
	Foundation.
\end{acknowledgments}

\appendix
\section{Effect of the Coordinate Rotation}\label{sec:app}

To illustrate the effect our rotation and reflection changes to \consf{HyperPipe} have made on the hyperparameter posteriors, we compare the results of our primary 6-factor inference with a 10\% buffer factor described in Sec. \ref{sec:result} (the ``rotated run'') to an identical analysis that did not apply these two modifications (the ``unrotated run''), retaining the $\gamma_k$ coordinates and traditional associated parameter bounds throughout the analysis. 

\begin{figure*}[t!]
	\centering
	\includegraphics[width=.45\textwidth]{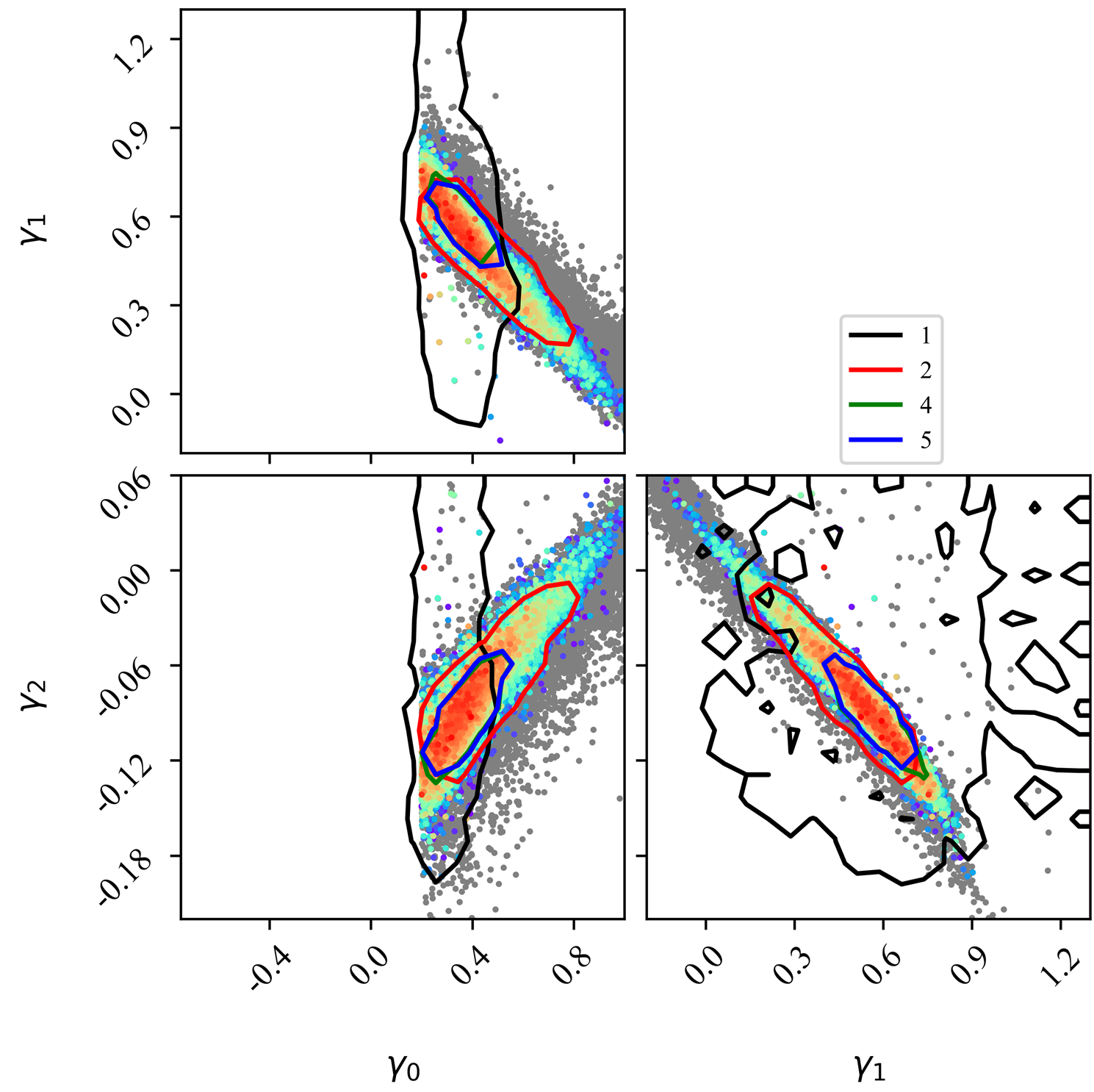}
	\includegraphics[width=.45\textwidth]{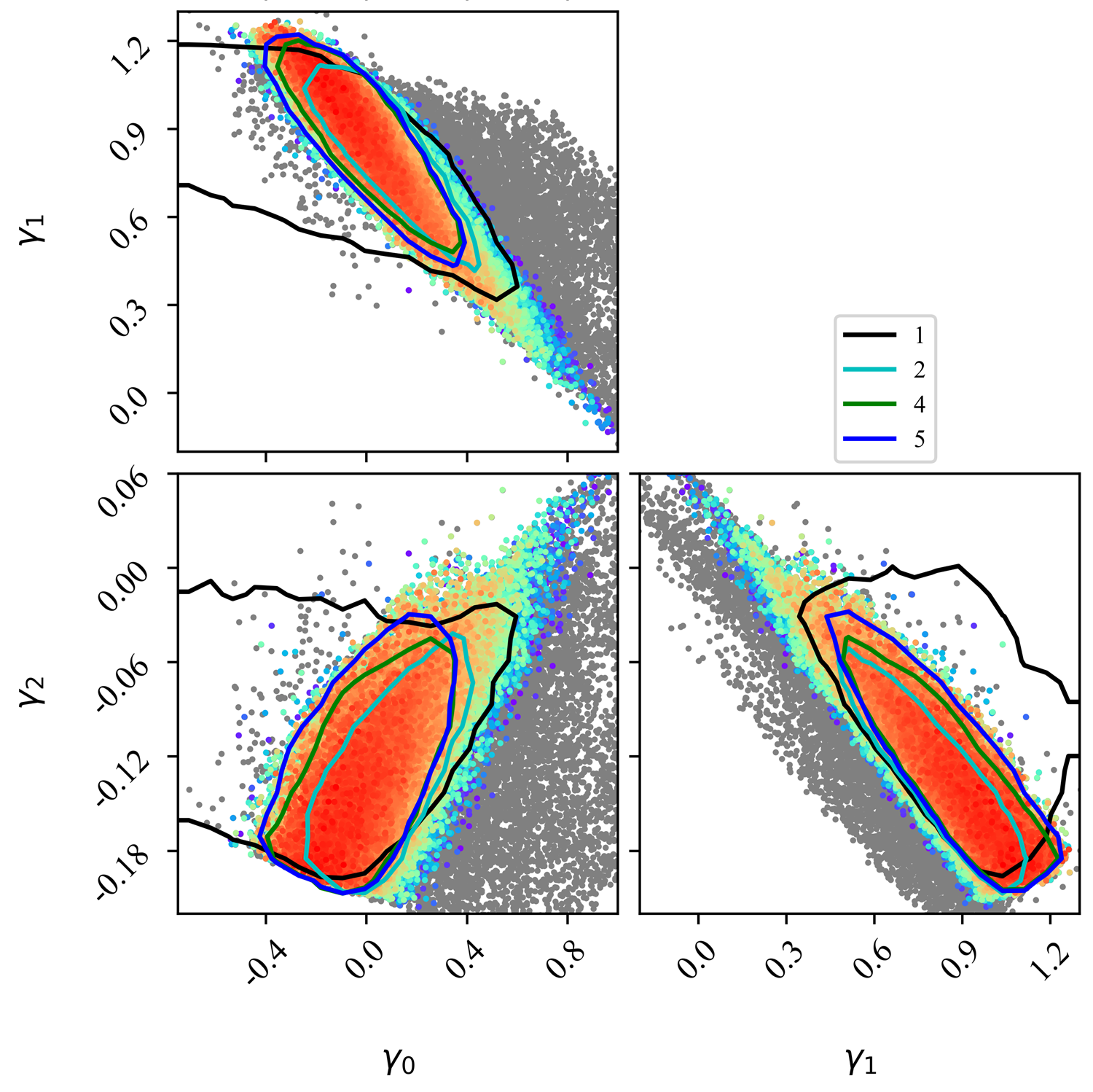}
	\caption{Corner plot comparison of the posterior evolution for three of the $\Gamma$-spectral EOS hyperparameters in the same 6-factor \consf{HyperPipe} inference, without (\textit{left}) and with (\textit{right}) the coordinate transformation and reflection. Iterations 1, 2, 4 and 5 are shown. The unrotated analysis slowly converges on a small region, while the rotated analysis quickly jumps to a small region before expanding. The axis ranges are the same in both plots to aid comparison. Several of the $\gamma_k$ and $r'_i$ parameter bounds are visible in both panels, truncating the posteriors. \label{fig:gamma_comp}}
\end{figure*}

For the initial sample set (iteration 0), the results between runs are identical. However, the first posteriors generated follow dramatically different behaviors, as shown in Fig. \ref{fig:gamma_comp}, which plots sequential posteriors from each run together on corner plots for some of the EOS parameters. As shown in the left-hand panel, the unrotated run first fits a broad posterior that is nearly entirely unphysical, since it is sampling over the Cartesian volume $\scC$. The algorithm then slowly converges around a small physical space in about 4 more iterations. On the other hand, the rotated run, as shown in the right-hand panel of the figure, samples its first posterior in the rotated hypercube $\scC'$ and therefore immediately finds better coverage of the physical space, honing in on a small region of the physical volume on the next iteration. Subsequently, the algorithm begins broadening its fit as \consf{HyperPipe}'s exploratory dithering phase finds more high-likelihood EOS points, attaining a more comprehensive sampling of the posterior region.

\begin{figure*}[t]
	\centering
	\includegraphics[width=.55\textwidth]{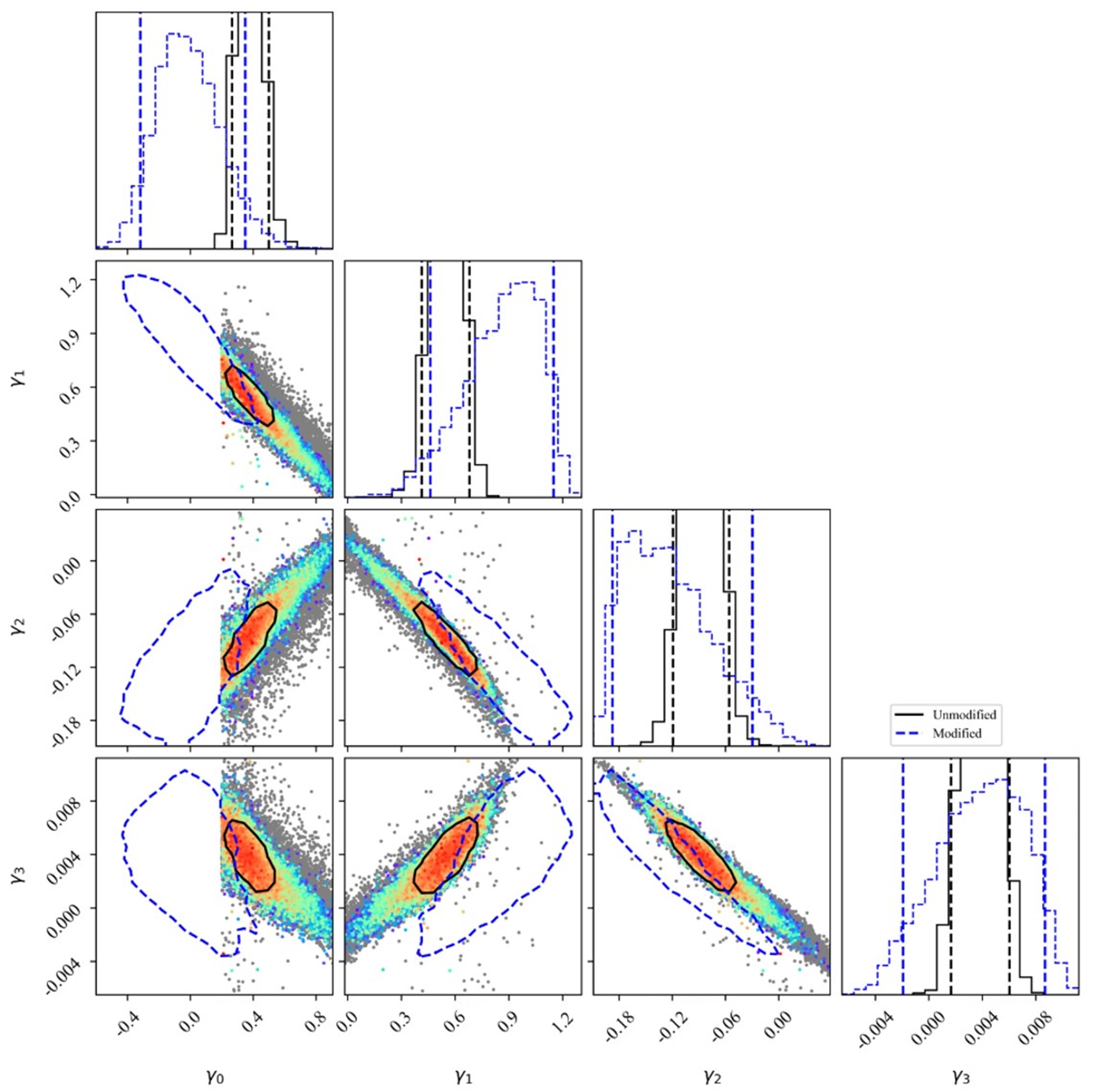}
	\includegraphics[width=.44\textwidth]{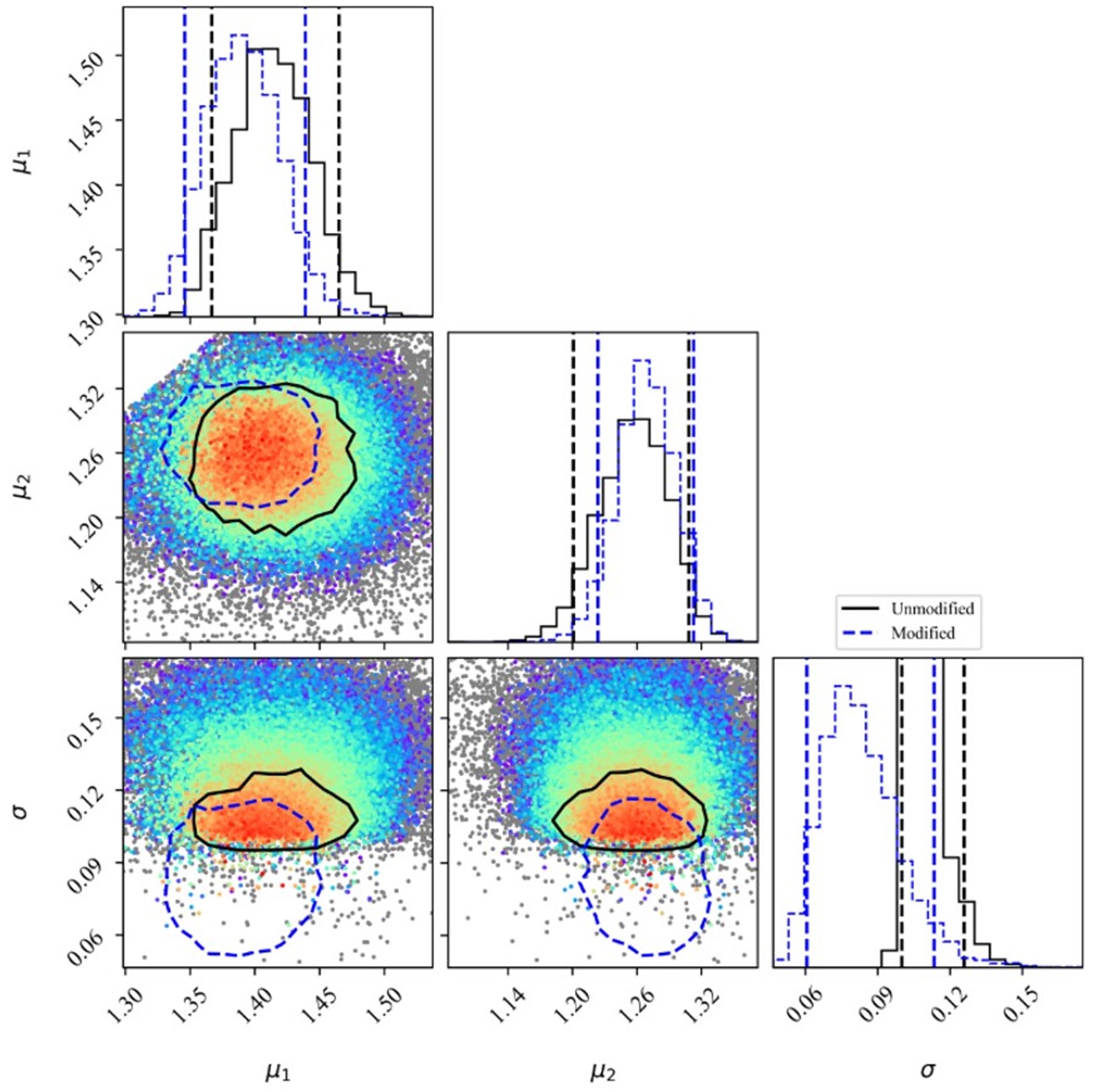}
	\caption{Corner plot of the 6-factor \consf{HyperPipe} inference analysis, without the coordinate transformation and reflection, after 6 iterations. The corresponding posteriors from the analysis using the transformation and reflection modifications are overlaid as dashed blue lines. The Cartesian cutoff of $\gamma_0 \ge 0.2$ is clearly visible on the left edge of the EOS parameter plots, as is the size difference between the posterior regions. \label{fig:rot_comp}}
\end{figure*}

The final 90\% credible posteriors for both runs are shown in Fig. \ref{fig:rot_comp} over top the accumulated samples of the unrotated run, with the unrotated posterior shown in solid black and the rotated posterior in dashed blue. 
These two posteriors differ in both size and location in parameter space, since the rotated run explores the (tightly-correlated) parameter space more efficiently and discovers more high-likelihood points, producing a large posterior displaced from that of the non-rotated analysis.
The rotated algorithm can find all these physically valid points because the rotated $\scC'$ space is aligned with the physical volume, enabling \consf{HyperPipe} to sample along it, whereas the Cartesian coordinate space essentially slices just a narrow cross-section of the physical volume.

The population hyperparameters are less strongly affected by these changes, but $\mu_1$ still shifts towards a smaller mean value in the rotated run and $\sigma$ is able to be sampled in its full parameter range; it is unclear why the sampling for $\sigma$ became truncated at $\sigma \approx 0.1\,\Msun$ in the unrotated run.

The parameter bounds in both $\scC$ and $\scC'$ are also clearly visible in both Figures \ref{fig:gamma_comp} and \ref{fig:rot_comp}, showing the clear truncations imposed on the posteriors in both runs, with the unrotated run enforcing $\gamma_0 \ge 0.2$ and the rotated run's posterior railing to one corner of the 10\%-buffered rotated hypercube. 
These artificial limits on the posterior demonstrate why the 400\% buffer on the rotated coordinate bounds explored in the main text should be the preferred bounds in future work using this EOS model family.

We conclude that using the rotated coordinate space to sample the EOS posteriors gives \consf{HyperPipe} vastly superior efficiency during early iterations, allowing it to quickly converge on a high-likelihood posterior. Reflecting puffed points also appears to prevent extremely unphysical samples from crashing the marginalization codes; the corresponding increase in the number of samples filling out the allowed space is easy to observe in the right-hand panel of Fig. \ref{fig:gamma_comp}. These factors combined significantly reduce the amount of computational time wasted on useless points and increase the density of valid sampled points, producing better-resolved posteriors for the EOS hyperparameters.

\bibliography{references}

\end{document}